\documentclass[a4paper,twoside,11pt]{article}
\usepackage[utf8]{inputenc}
\usepackage[T2A]{fontenc}
\usepackage[english]{babel}
\usepackage{fancyhdr}
\usepackage{newprog1e}
\usepackage{amsfonts,amsmath,amssymb,amsthm}
\usepackage{graphicx}
\usepackage{listings}
\usepackage{tikz}
\usetikzlibrary{positioning, arrows}
\usepackage{algorithm}
\usepackage{algorithmicx}
\usepackage{algpseudocode}
\usepackage{makecell}
\usepackage[nocolor]{drawstack}
\usepackage{adjustbox}
\usepackage{booktabs}
\usepackage{multirow}
\usepackage{colortbl}
\usepackage{boldline}
\usepackage{subfig}
\usepackage[labelsep=space, textformat=period, labelfont=bf]{caption}
\addto\captionsenglish{\renewcommand{\figurename}{Fig.}}
\usepackage{pifont}
\newcommand{\xmark}{\ding{55}}%
\newfloat{listing}{thp}{lol}
\floatname{listing}{Listing}
\usepackage{indentfirst}

\usepackage[hyphens]{url}
\usepackage[unicode, hidelinks, bookmarksnumbered]{hyperref}
\usepackage[hyphenbreaks]{breakurl}
\usepackage{bookmark}
\usepackage[autostyle]{csquotes}
\usepackage[
    backend=biber,
    bibencoding=utf8,
    style=ieee,
    sortcites=true,
    citestyle=numeric-comp,
    natbib=true,
    url=true,
    doi=true,
    eprint=false
]{biblatex}
\defbibheading{bibliography}[\bibname]{
\pdfbookmark{REFERENCES}{REFERENCES}
\begin{center}
\hspace{1em}{REFERENCES}
\end{center}
}
\usepackage{titlesec}
\titleformat{name=\section}
{\filcenter}
{\thesection.}
{1em}
{}
\titleformat{name=\subsection}
{\filcenter\itshape}
{\slshape{\thesubsection}.}
{1em}
{}
\titleformat{name=\subsubsection}
{\filcenter\itshape}
{\slshape{\thesubsubsection}.}
{1em}
{}

\numberwithin{equation}{section}

\journalvolume{47}
\journalnumber{4}
\curyear{2021}
\authorlist{VISHNYAKOV, NURMUKHAMETOV}
\titlehead{SURVEY OF METHODS FOR AUTOMATED CODE-REUSE EXPLOIT GENERATION}
\headerdef

\udk{000.000.0}
\rubrika{}
\dateinput{November 11, 2020; revised December 5, 2020; accepted December 20, 2020}
\DOI{10.1134/S0361768821040071}

\rusabstr{%
\textbf{Abstract}---This paper provides a survey of methods and tools for automated code-reuse exploit
generation. Such exploits use code that is already contained in a vulnerable program. The code-reuse
approach allows one to exploit vulnerabilities in the presence of operating system protection that
prohibits data memory execution. This paper contains a description
of various code-reuse methods: return-to-libc attack, return-oriented programming, jump-oriented
programming, and others. We define fundamental terms: gadget, gadget frame, gadget
catalog. Moreover, we show that, in fact, a gadget is an instruction, and a set of gadgets defines
a virtual machine. We can reduce an exploit creation problem to code generation for this virtual
machine. Each particular executable file defines a virtual machine instruction set. We provide
a survey of methods for gadgets searching and determining their semantics (creating a gadget
catalog). These methods allow one to get the virtual machine instruction set. If a set of gadgets is
Turing-complete, then a compiler can use a gadget catalog as a target architecture. However, some
instructions can be absent. Hence we discuss several approaches to replace missing instructions
with multiple gadgets. An exploit generation tool can chain gadgets by pattern searching (regular
expressions) or considering gadgets semantics. Furthermore, some chaining methods
use genetic algorithms, while others use SMT-solvers. We compare existing open-source tools
and propose a testing system rop-benchmark that can be used to verify whether a generated chain
successfully opens a shell.%
}

\author{
  {\bfseries A.~V.~Vishnyakov$^{a,*}$ and A.~R.~Nurmukhametov$^{a,**}$}
\\ {\itshape $^a$ Ivannikov Institute for System Programming, Russian Academy of Sciences,}
\\ {\itshape Moscow, }{\slshape 109004 }{\itshape Russia}
\\ {\itshape *e-mail: vishnya@ispras.ru }
\\ {\itshape **e-mail: nurmukhametov@ispras.ru }}
\title{Survey of Methods for Automated Code-Reuse Exploit Generation}
\date{}

\begin{document}

\maketitle
\setcounter{page}{1}
%%%%%%%%%%%%%%%%%%%%%%%%%%%%%%%%%%%%%%%%%%%%%%%%%%%%%%%
\section{INTRODUCTION}
%%%%%%%%%%%%%%%%%%%%%%%%%%%%%%%%%%%%%%%%%%%%%%%%%%%%%%%

Modern software is known to contain bugs. Some researchers consider such errors as
inevitable. However, not all errors can be used to harm. Exploitable errors are
called vulnerabilities. When exploited, the vulnerability causes serious consequences such as
money losses, degradations of communication, compromise of cryptographic
keys~\cite{heartbleed}. With the development of the Internet of things,
one can exploit things that surround us daily, such as kettles,
refrigerators, and shower systems.
Medical equipment safety issues are crucial. Halperin et al.~\cite{halperin08} showed that it is
possible to exploit implanted heart defibrillators.

Along with the security development lifecycle, methods to detect various software
defects are also improving. In response to the improvement of protection methods
against vulnerabilities exploitation, new methods are being developed to bypass and
exploit them. Hence, it is necessary to know and understand the principles of both
software protection and attacks. 
Moreover, vendors and software developers can require a proof of concept (exploit)
to prioritize a vulnerability fix.

The stack buffer overflow is likely to be one of the most exploited software
defects~\cite{cwe-top25} because it is easy to use it for the control-flow hijacking.
In simplest case of a total lack of protection, exploitation goes as follows.
The return address located on stack above the local buffer is overwritten with a
controlled value. This value points back to the buffer that contains the code that the
attacker wants to execute.

The DEP protection has appeared to counter the code execution on stack.
DEP prohibits writing to code regions and execution in stack and heap process memory.
This protection puts an end to the code injection into process
memory. The attackers were restricted in executing just the
code available in process memory. In response to DEP ubiquity, code-reuse attacks
began to be developed rapidly. The first one was a return-to-libc attack~\cite{return-into-libc}.
The return address is replaced by the address of the function to be called, followed
by its arguments. Return-oriented programming (ROP)~\cite{shacham07, schwartz11, roemer12} is a generalization of this
technique. In return-oriented programming, one uses gadgets instead of functions.
A \textit{gadget} is a short instruction sequence ending with a return instruction. Gadgets are
chained together so that they sequentially transfer control to one another and carry out a
malicious payload. Shacham~\cite{shacham07} defined the term gadget and introduced the first gadget
catalog for the x86 instruction set. He also proved that this catalog is Turing-complete.
After that, the applicability of ROP was also shown for other
architectures:
ARM~\cite{kornau10, checkoway10, davi10, huang12, fraser17},
SPARC~\cite{buchanan08},
Atmel AVR~\cite{francillon08},
PowerPC~\cite{lindner09},
Z80~\cite{checkoway09},
MIPS~\cite{lindner09}.
Papers~\cite{bletsch11, checkoway10, chen11, sadeghi17}
showed that gadgets ending with not only return instructions could be used.

ROP can be
used as a steganographic method~\cite{lu14}. Ntantogian et al.~\cite{ntantogian19} proposed using code-reuse
methods to hide malicious functionality from detecting by antivirus tools, while Mu et
al.~\cite{mu17} suggested using ROP to obfuscate the code. Code-reuse methods
allow for backdoors to be inserted into software~\cite{bosman14, borrello19}.

With the development of the code-reuse methods, tools were also being developed,
helping the attacker construct such attacks. At first, this process was almost manual, but
over time it gradually became automated. At the moment, the literature presents a set of
approaches to automated construction of code-reuse exploits~\cite{hund09,
buchanan08, schwartz11, chen11, roemer12, huang12, quynh13, ding14, ouyang15,
follner16, fraser17, milanov18, mosier19}.
For some of them, even the tools are available~\cite{pshape, roper, monapy,
ropgadget, ropgenerator, angrop, ropper, ropc, ropc-llvm, pyrop, sqlabropchain,
ropilicious, exrop}.

This work aims to provide a detailed study of available methods and tools for automated code-reuse
exploit generation to determine their strengths and weaknesses and identify
future directions for the research.

In addition to practical use, methods and tools considered in the paper may be of
scientific interest. The task of automated exploit generation is
to translate some exploit description into code for the virtual machine instruction
set architecture, implicitly set by the memory state of the exploitable process.
Gadgets in the process memory are like instructions.
What is more, the exploited executable file provides an instruction set that is
not known a priori.
To learn this instruction set, one needs to find all the gadgets and determine their
functionality (semantics). As a result, one creates a catalog of gadgets that
describes their semantics. A gadget catalog is an input data for the tool that generates
exploits. The exploit generation tool should consider that a set of gadgets,
unlike processor instructions, may lack some instructions, while others may have non-trivial
side effects. It complicates the development of the tools for automated
generation of ROP exploits.

The paper has the following structure.
Sections~\ref{sec:background}--\ref{sec:dop} provide a survey of attacks and defense mechanisms.
Section~\ref{sec:generation-scheme} describes a general scheme for code-reuse exploit generation.
Section~\ref{sec:catalogue} introduces a definition of the \textit{gadget catalog}.
Section~\ref{sec:gadget-search} describes approaches to finding gadgets.
Section~\ref{sec:semantic-definition} provides methods to determine the gadget semantics.
Section~\ref{sec:chain-generation} reviews methods for gadget chains generation.
Section~\ref{sec:badchars} reveals the problem of accounting restricted symbols in chains.
Section~\ref{sec:evaluation} presents an experimental comparison of open-source tools
done by the specially developed rop-benchmark testing system~\cite{ropbenchmark}.
The last Section~\ref{sec:conclusion} discusses the problems of the existing methods and
identifies further research directions.

%%%%%%%%%%%%%%%%%%%%%%%%%%%%%%%%%%%%%%%%%%%%%%%%%%%%%%%
\section{BACKGROUND}
\label{sec:background}
%%%%%%%%%%%%%%%%%%%%%%%%%%%%%%%%%%%%%%%%%%%%%%%%%%%%%%%

%%%%%%%%%%%%%%%%%%%%%%%%%%%%%%%%%%%%%%%%%%%%%%%%%%%%%%%
\subsection{Data Execution Prevention}
\label{sec:dep}
%%%%%%%%%%%%%%%%%%%%%%%%%%%%%%%%%%%%%%%%%%%%%%%%%%%%%%%

Data Execution Prevention is an operating system protection that prohibits
execution of memory pages marked as ``data''. A memory page can be
simultaneously accessible either for writing or for execution, but not both, which
precisely reflected in the name of the OpenBSD W\^{}X security policy~\cite{wxorx} (\textbf{W}rite XOR
e\textbf{X}ecute). On Windows, this defense mechanism is called DEP (Data Execution
Prevention)~\cite{dep}. Linux~\cite{vandeven04} and Mac OS X have similar protective mechanisms. The
protective mechanism is implemented in the hardware using a special NX-bit (\textbf{N}o
e\textbf{X}ecute), which marks the pages inaccessible for execution. If the processor lacks the
hardware support for the NX-bit, then that mechanism is emulated by the software.

In classic exploitation of stack buffer overflow~\cite{tanenbaum2015}, the attacker injects a malicious code
into the buffer and transfers the control onto that code. The protection does not allow one
to execute the injected code as its location is the stack marked as ``data''.

%%%%%%%%%%%%%%%%%%%%%%%%%%%%%%%%%%%%%%%%
%\vspace{0.25cm}
\subsection{Code-Reuse Attacks}
%%%%%%%%%%%%%%%%%%%%%%%%%%%%%%%%%%%%%%%%

Code-reuse attacks have appeared to bypass the protection that prevents data execution.
The idea is not to inject a malicious code but to reuse the code already presented
in the program and libraries to implement the functionality of the malicious
code. The stack buffer overflow vulnerability or the ability to
write an arbitrary value to an arbitrary memory location (write-what-where~\cite{cwe-123})
allows one to replace the return address with the address of
some code from the program address space. Thus, after returning from the
function, the control is transferred to this code.

%%%%%%%%%%%%%%%%%%%%%%%%%%%%%%%%%%%%%%%%
\subsection{Return-to-Library Attack}
%%%%%%%%%%%%%%%%%%%%%%%%%%%%%%%%%%%%%%%%

Alexander Peslyak was the first to show that exploitation is possible even with
a non-executable stack and proposed a return to library attack (return-to-libc)~\cite{return-into-libc}.
The attacker substitutes the return address with
some library function address and places its arguments up the stack. For example, the
attacker may call \verb|system("/bin/sh")| from the standard library \verb|libc|.
Thus, the attacker may open the operating system shell.

%%%%%%%%%%%%%%%%%%%%%%%%%%%%%%%%%%%%%%%%
\subsection{Address Space Layout Randomization}
\label{sec:aslr}
%%%%%%%%%%%%%%%%%%%%%%%%%%%%%%%%%%%%%%%%

An address space layout randomization (ASLR)~\cite{aslr} is a protective
mechanism of the operating system that loads memory segments at different base
addresses for each program run. This protection makes it difficult to conduct a return-to-library
attack (return-to-libc), as the base address of the library \verb|libc| is
random, and the \verb|system| function address is unknown before the program
loading. However, for compatibility with ASLR, the program must be compiled into a
position-independent code~\cite{bhatkar03}, which is not always held. For example, in Linux, the
base addresses of dynamic libraries, stack and heap are randomized, while the base
address of the program image often remains constant~\cite{fedotov16}.

If the library base address is random, but the program image is not, then the
attacker can call the imported function through the procedure linkage table PLT~\cite{plt},
which contains a code for calling library functions. The return-to-plt attack is a
modification of the return-to-library attack and consists of replacing the return
address with the address of the code from PLT that calls the function from the
dynamic library.

%%%%%%%%%%%%%%%%%%%%%%%%%%%%%%%%%%%%%%%%
\section{RETURN-ORIENTED PROGRAMMING}
%%%%%%%%%%%%%%%%%%%%%%%%%%%%%%%%%%%%%%%%

Shacham~\cite{shacham07} suggested the term return-oriented programming (ROP). ROP is an
effective method to bypass data execution prevention (DEP~\cite{dep}, W\^{}X~\cite{wxorx}). In a
certain sense, this approach is the generalization of the return-to-library attack.
However, the malicious payload is implemented not by calling one function, but is
formed from several code pieces already present in the program, which are called
\textit{gadgets}. A gadget is an instruction sequence ending with a control transfer
instruction. Each gadget modifies the state of registers and memory. For instance, it
adds the values of two registers and writes the result to the third one.
Having studied all the gadgets available in the program, the attacker links them into chains in
which the gadgets sequentially transfer the control to each other. 
Such a chain of gadgets carries out the total malicious payload.
With a sufficient number of
gadgets, the attacker can form a Turing-complete set to allow arbitrary
computation~\cite{shacham07}. It is worth noticing that ROP is also useful when partial
randomization of the address space is present. In this case, gadgets from non-randomized
memory areas are used.

%{{{ примеры гаджетов
\begin{table}[t]
  \caption{Example of x86 gadgets}
  \small
  \begin{tabular}{l l}
    \toprule
    \texttt{mov eax, ebx ; ret}
      & \makecell[cl]{Copying  \texttt{ebx} register value \\
                      into \texttt{eax} register.} \\
    \texttt{pop ecx ; ret}
      & \makecell[cl]{Loading the value from \\
                      stack on \texttt{ecx} register.} \\
    \texttt{add eax, ebx ; ret}
      & \makecell[cl]{Adding \texttt{ebx} register value \\
                      to \texttt{eax} register.} \\
    \bottomrule
  \end{tabular}
  \label{tbl:x86_gadgets}
\end{table}
%}}}

For clarity, Table~\ref{tbl:x86_gadgets} demonstrates the assembly
code\footnote{Hereafter, we will use Intel syntax for x86 assembler.}
of the three gadgets for x86. Each of the gadgets ends with
a \verb|ret| instruction, which allows transferring control to the next gadget via the address
placed on stack.

The x86 architecture is CISC. The x86 instructions are not fixed in length,
and each instruction can execute several other low-level commands.
The number of commands is enormous, and their encoding is so
tight that almost any sequence of bytes is the correct
instruction. Besides, due to the different command lengths (from 1 byte to
15), the x86 architecture does not require instructions alignment. From the
ROP point of view, it means the following. The set of gadgets in the
program is not limited only to compiler-generated instructions.
It enlarges with the instructions not presented in the
original program but received upon access to the middle of other commands.
Here is an illustrating example~\cite{salwan14}:
%{{{ невыровненный доступ к~инструкциям (пример)
\begin{lstlisting}[
    mathescape,
    basicstyle=\scriptsize\ttfamily,
    columns=fullflexible,
    keepspaces=true
]
f7c7070000000f9545c3 $\rightarrow$ test edi, 0x7 ;
                        setnz BYTE PTR [ebp-0x3d]
  c7070000000f9545c3 $\rightarrow$ mov DWORD PTR [edi], 0xf000000 ;
                        xchg ebp, eax ; inc ebp ; ret
\end{lstlisting}

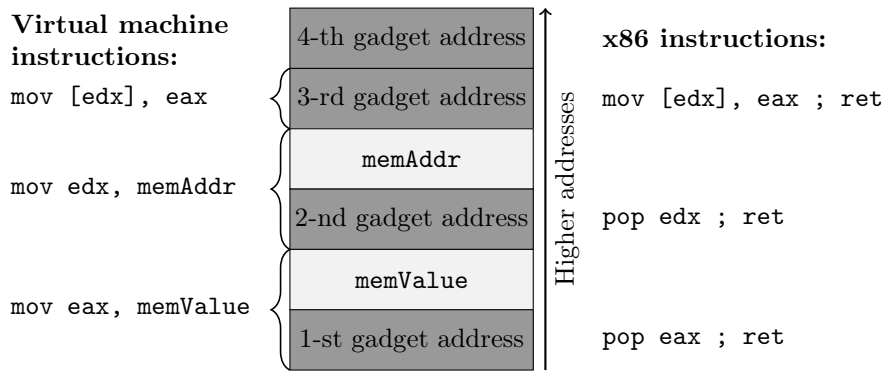
\begin{figure*}[t]
  \centering
  \small
  \begin{tikzpicture}[scale=0.8]
    \node[align=left, anchor=west] at (0,0) {\textbf{Virtual machine} \\
                                             \textbf{instructions:}};
    \node[align=left, anchor=west] at (0,-0.95) {\texttt{mov [edx], eax}};
    \node[align=left, anchor=west] at (0,-2.45) {\texttt{mov edx, memAddr}};
    \node[align=left, anchor=west] at (0,-4.45) {\texttt{mov eax, memValue}};
    \node at (0,-5.35) {};
  \end{tikzpicture}
  \trimbox{3.2cm 0cm 0cm 0cm}{
  \begin{tikzpicture}[scale=0.8]
    \bcell{4-th gadget address} \cellcom{~~~~\textbf{x86 instructions:}}
    \startframe
    \bcell{3-rd gadget address} \cellcom{~~~~\texttt{mov [edx], eax ; ret}}
    \finishframe{}
    \startframe
    \cell{\texttt{memAddr}}
    \bcell{2-nd gadget address} \cellcom{~~~~\texttt{pop edx ; ret}}
    \finishframe{}
    \startframe
    \cell{\texttt{memValue}}
    \bcell{1-st gadget address} \cellcom{~~~~\texttt{pop eax ; ret}}
    \finishframe{}
    \draw[->, thick] (2.2, -6.5) -- (2.2, -0.5)
        node[midway, below, sloped] {Higher addresses};
  \end{tikzpicture}%
  }
  \caption{A ROP Chain, storing \texttt{memValue} to \texttt{memAddr}}
  \label{fig:rop-machine}
\end{figure*}
%}}}

An executable file essentially defines the set of gadgets that can be used to compose a ROP chain.
Furthermore, for another executable file, the ROP chain
has to be reassembled anew. The ROP chain can be considered as a program
for some virtual machine defined by an executable file~\cite{dullien17}. The stack pointer acts
as a program counter for this virtual machine. The operation codes (gadget
addresses) and their operands are located on stack. Graziano et al.~\cite{graziano16} even
proposed a tool for translating ROP chains into a regular x86 program. Figure~\ref{fig:rop-machine} shows
an example of a gadget chain located on stack that stores \verb|memValue| to \verb|memAddr|.
Opcodes (gadget addresses) are located from the return address on stack and are
shaded in dark gray. The operands \verb|memValue| and \verb|memAddr| are shaded in light gray.
Curly brackets denote virtual machine instructions (opcode and its operands). The
actual gadget instructions on x86 are given on the right. At the beginning of the
chain, where in normal execution the function return address is located, we place the
opcode~-- the address of the first gadget. Then the \verb|memValue| operand is located, which
the first gadget loads into the \verb|eax| register. Then follows the address of the second
gadget to which the first gadget transfers control via the ret instruction, and so on.

Return instructions on x86 are encoded as: \verb|c3|, \verb|c2**|, \verb|cb|, \verb|ca**| (where any bytes can
be instead of stars). Encoding the return instruction in such a way results in many gadgets
in x86 code. Even relatively small binary files contain gadgets that are practically
applicable from the attackers' point of view. Schwartz et al.~\cite{schwartz11, schwartz11_update} provide statistics that
among the programs larger than 100KB, about 80\% contain sets of gadgets that allow
one to call any function from the library which is dynamically linked with a vulnerable
application.

Subsequently, the use of ROP was successfully demonstrated for other
architectures: SPARC~\cite{buchanan08}, Z80~\cite{checkoway09} (Harvard architecture voting machine of the late 80s),
ARM~\cite{kornau10, checkoway10, davi10, huang12, fraser17, weidler18}.
The above-mentioned works showed that on RISC architectures it
was possible to construct a set of gadgets both serviceable and Turing-complete. RISC
architectures are often characterized by fixed commands length, requirement to
align instructions on their size, and simplified access to memory (only store and load
instructions access to memory). The instructions alignment, compared to x86,
forces attacker to find gadget that the program code originally contained. These
gadgets are usually valid function epilogues.

%%%%%%%%%%%%%%%%%%%%%%%%%%%%%%%%%%%%%%%%
\subsection{Gadget Frame}
\label{sec:gadget-frame}
%%%%%%%%%%%%%%%%%%%%%%%%%%%%%%%%%%%%%%%%

\begin{figure}[t]\centering
\small
\trimbox{3.8cm 0cm 0cm 0cm}{
\begin{drawstack}[scale=0.75]
    \startframe
    \padding{2}{}
    \cell{Next gadget}
    \cell{``Loaded'' \texttt{eax}}
    \finishframe{}
    \bcell{\texttt{pop eax ; ret 8}}
\end{drawstack}}
\hbox{}
\begin{tikzpicture}[scale=0.8]
    \draw[->, thick] (0, 0) -- (0, 7)
        node[midway, below, sloped] {Higher addresses};
\end{tikzpicture}
\caption{\texttt{pop eax ; ret 8} gadget frame}
\label{fig:gadget-frame}
\end{figure}

In order to place a ROP chain on stack, it is convenient to introduce the concept of a
\textit{gadget frame}~\cite{vishnyakov18} similar to the x86 stack frame.
A chain of gadgets is assembled from
the frames. The gadget frame contains values of gadget parameters (for
instance, the value loaded onto the register from the stack) and the address of
the next gadget. The beginning of the frame is determined by the value of
the stack pointer before executing the first gadget instruction. In Figure~\ref{fig:gadget-frame},
curly brackets denote the borders of the pop eax; ret 8 gadget frame. The gadget
loads the value from the stack into \verb|eax| at offset 0 from the beginning of the frame.
The size of the gadget frame is \texttt{FrameSize = 16}, and the next gadget address is located
at offset 4 from the beginning of the frame (\texttt{NextAddr = [esp + 4]}).

%%%%%%%%%%%%%%%%%%%%%%%%%%%%%%%%%%%%%%%%
\subsection{Returning to Randomized Library Attack}
%%%%%%%%%%%%%%%%%%%%%%%%%%%%%%%%%%%%%%%%

Roglia et al.~\cite{roglia09} showed how to call a function from the library with ROP,
even though the library base address is randomized (the vulnerable program
image base considered not to be randomized). Linux stores
the addresses of imported functions in the section \verb|.plt.got|~\cite{plt} to
perform dynamic linkage (Windows has a similar
mechanism via the Import Address Table~\cite{import-address-table}).
The attacker can use this information to calculate the addresses of
the remaining functions from dynamically linked libraries.
Suppose that \verb|.plt.got| contains the address of the imported \verb|open| function from
\verb|libc|. Then the address of function \verb|system| can be calculated with the following
formula: $$system = open + (offset(system) - offset(open))$$
where $offset(s)$ function returns the offset of $s$ function relatively to the library base
address. ASLR randomizes the base address of the library load, while the
offset value of \verb|system| function relative to \verb|open| inside the library ($offset(system) -offset(open)$)
remains constant and is known to the attacker in advance.

The attacker creates a ROP chain which loads the address of \verb|open| function
from \verb|.plt.got|, add the previously known offset of \verb|system| relative to \verb|open| to the loaded
address, and transfer the control to the calculated address, i.e., to \verb|system| function.
The attacker can also make a chain that adds to the address of \verb|open| function a
necessary offset in the \verb|.plt.got|
memory and call the function by its address from that memory. If it is necessary to call
the imported function, then a ROP chain can be made up that calls the function by its
address from \verb|.plt.got|, or one can perform the return-to-plt attack (Section~\ref{sec:aslr}), where the
code in PLT calls this function.

It is worth noticing that Linux uses the lazy binding mechanism. Initially, the
addresses of the stub functions are written in \verb|.plt.got| instead of the
addresses of the imported functions.
When the imported function is called for the first time, the stub function
dynamically binds it and write its virtual address in \verb|.plt.got|. Thus, the address of
\verb|system| function should be calculated based on the address of the function, already called
from \verb|libc| at the time of exploitation, i.e., which address is already recorded
in \verb|.plt.got|.

To protect \verb|.plt.got| from being overwritten, there is \verb|LD_BIND_NOW| flag, which disables
lazy binding and tells the loader to bind all imported functions immediately~\cite{ld}.
However, reading from \verb|.plt.got| is still possible and one can calculate the address of
\verb|system| based on the address of any imported function.

Kirsch et al.~\cite{kirsch17} showed that even with turned on protections, the dynamic loader
(POSIX) leaves in the program writable pointers to the functions that are called when exiting
the program. The attacker can overwrite these pointers so that
malicious code is executed during the program exit.

%%%%%%%%%%%%%%%%%%%%%%%%%%%%%%%%%%%%%%%%
\subsection{Using Gadgets from a Randomized Library}
%%%%%%%%%%%%%%%%%%%%%%%%%%%%%%%%%%%%%%%%

The gadgets in the vulnerable executable file are not always sufficient to implement the
malicious payload. For example, there may be no gadgets to load function arguments
passed through registers. Ward et al.~\cite{ward19} proposed a method that allows using gadgets
from dynamically linked libraries, whose base addresses are randomized. It is assumed that
the base address of the vulnerable executable file is not therewith randomized.

The idea is based on the ability to partially overwrite pointers from the global offset
table (GOT~\cite{plt}). Such overwriting can be performed, for example, if a write-what-where~\cite{cwe-123}
condition is present. The table contains the values of pointers to code in the
process memory (usually located in libraries). By changing the last byte of the pointer
value one can address the code within the memory paragraph around this pointer. For
example, if the pointer value is 0xdeadbeef, then an address range 0xdeadbe00--0xdeadbeff
is available for addressing. Address space randomization, operating on the
level of changing virtual memory tables, only changes the high-order bytes of addresses.
Thus, the code located on one page does not change its low-order bytes. It follows that
rewriting the low-order byte of the pointer to the code allows positionally independent
addressing of the code within a memory paragraph of size $2^8 = 256$ bytes.

After the low bytes of the pointers in the Global Offset Table (GOT) are corrected,
control should be transferred to them. To achieve this, one fills the stack with pointers
to the records of the procedure linkage table (PLT~\cite{plt}), which indirectly transfers
control to the addresses recorded in the corresponding cells of the GOT table.
It is worth noticing that one can only use records of those functions whose
addresses were filled
by a dynamic loader (i.e., those called at least once before
exploitation). One can call gadgets that lie within the same memory
paragraph from the beginning of the function. Thus, one can call gadgets from a
randomized library.

%%%%%%%%%%%%%%%%%%%%%%%%%%%%%%%%%%%%%%%%
\subsection{Stack Pivot}
\label{sec:stackpivot}
%%%%%%%%%%%%%%%%%%%%%%%%%%%%%%%%%%%%%%%%

Dino Dai Zove~\cite{dai10} introduced a \textit{trampoline gadget} (\textit{Stack Pivot}), which
can be used when exploiting stack or heap buffer overflows as an intermediate
link. The stack pivot moves the stack pointer to the beginning of the ROP chain and
thereby transfers control to it. Stack Pivots are as follows:
\begin{itemize}
  \item \texttt{mov esp, eax ; ret}
  \item \texttt{xchg eax, esp ; ret}
  \item \texttt{add esp, <constant> ; ret}
  \item \texttt{add esp, eax ; ret}
\end{itemize}

The attacker can replace the function pointer with the address of the stack
pivot, for instance, using a fake virtual functions table formed on heap. Instead of
calling the function, the stack pivot moves the stack pointer to the beginning of
the ROP chain.

%%%%%%%%%%%%%%%%%%%%%%%%%%%%%%%%%%%%%%%%
\subsection{Canary Bypass}
%%%%%%%%%%%%%%%%%%%%%%%%%%%%%%%%%%%%%%%%

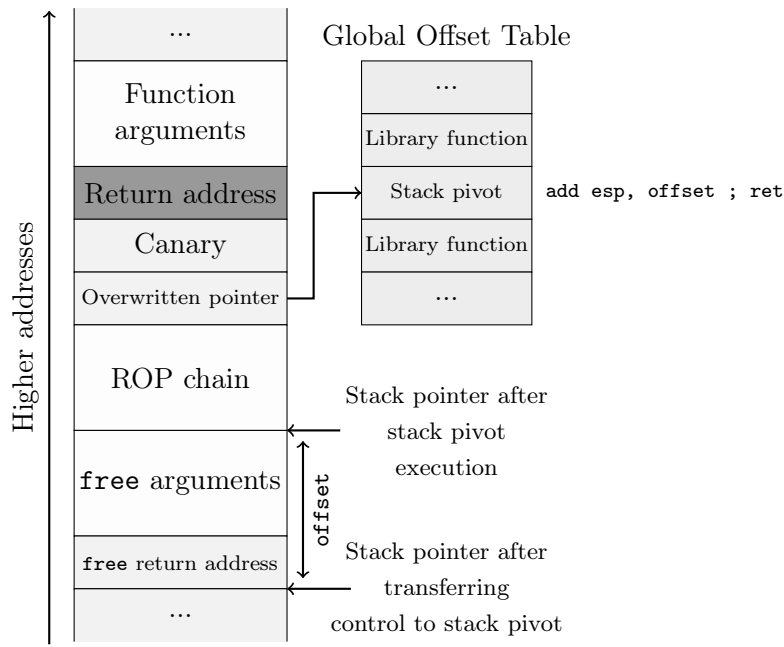
\begin{figure*}[t]\centering
\begin{tikzpicture}
    \draw[->, thick] (0, 0) -- (0, 8.4)
        node[midway, above, sloped] {Higher addresses};
\end{tikzpicture}
\hbox{}
  \begin{tikzpicture}[scale=0.7]
    \stacktop{}
    \padding{2}{}
    \bcell{Return address}
    \cell{Canary}
    \cell{\scriptsize Overwritten pointer}
    \padding{2}{ROP chain}
    \padding{2}{\texttt{free} arguments}
    \cell{\scriptsize \texttt{free} return address}
    \stackbottom{}

    \drawstruct{(5, 0)}
    \structcell{...}
    \structcell{\scriptsize Library function}
    \structcell{\scriptsize Stack pivot}
    \structcell{\scriptsize Library function}
    \structcell{...}

    \draw (9.1, -3) node{\scriptsize \texttt{add esp, offset ; ret}};
    \draw (5, -7.5) node[align=center] {\footnotesize Stack pointer after\\
                                        \footnotesize stack pivot \\
                                        \footnotesize execution};
    \draw (5, -10.5) node[align=center] {\footnotesize Stack pointer after\\
                                         \footnotesize transferring \\
                                         \footnotesize control to stack pivot};
    \draw (5, 0) node {Global Offset Table};
    \path [draw, ->, thick] (2, -5) -- (2.5, -5) |- (3.4, -3);
    \draw [->, thick] (3, -7.5) -- (2, -7.5);
    \draw [->, thick] (3.2, -10.5) -- (2, -10.5);
    \draw [<->, thick] (2.3, -10.3) -- (2.3, -7.7)
        node[midway, below, sloped] {\footnotesize \texttt{offset}};
    \draw (0, -1.5) node[align=center] {Function \\
                                      arguments};
  \end{tikzpicture}
  \caption{Stack canary bypass}
  \label{fig:canary}
\end{figure*}

The compiler uses canaries~\cite{cowan98} to protect against stack buffer overflow exploitation.
During the function call, the compiler inserts an arbitrary value just before
the return address on stack. This value is called a ``canary''.
The compiler prepends the return from the function by a code that checks the canary value.
If the value has
changed, the program crashes. Thus, it becomes impossible to place the ROP
chain starting from the return address and execute it because that overwrites
and changes the canary value.

Fedotov et al.~\cite{fedotov16} showed how to bypass the canary when the data execution prevention
is operating. The method can be applied if there is a write-what-where condition~\cite{cwe-123}:
\begin{enumerate}
  \item Buffer overflow causes overwriting of the pointer placed on stack.
  \item The attacker controls the value that is written by this pointer.
\end{enumerate}

Suppose that after overflow and before checking the canary, \verb|free| function is
called (Fig.~\ref{fig:canary}). Then, the attacker overwrites the pointer with the address of the
GOT table cell in which the address of \verb|free| function is stored. To the cell of \verb|free|
function in the GOT table one writes the address of the stack pivot which moves the
stack pointer and transfers control to the ROP chain located up the stack, but
before the canary. Thus, instead of calling \verb|free| function, control is transferred to
the stack pivot that, in turn, transfers control to the ROP chain. As this occurs,
the canary does not change, and verification of its value upon
return from the function is successful.

%%%%%%%%%%%%%%%%%%%%%%%%%%%%%%%%%%%%%%%%
\subsection{Disabling DEP and Transferring Control to a Regular Shellcode}
\label{sec:virtualprotect}
%%%%%%%%%%%%%%%%%%%%%%%%%%%%%%%%%%%%%%%%

A two-stage method for exploitation is common~\cite{dai10} with ROP as the first stage.
It is responsible for placing the second stage shellcode, disabling protections, and transferring control to the
injected shellcode. The second stage executes a regular shellcode which contains
the main malicious payload. Thus, it is possible to modify the malicious
payload by replacing just the second stage shellcode. The
following is a detailed description of both stages:
\begin{enumerate}
  \item \textbf{ROP stage.} An attacker places a shellcode on stack or
    writes it into memory with a ROP chain. Next, the attacker makes up a ROP
    chain that disables DEP: calling the \verb|mprotect| function~\cite{mprotect}
    (\verb|VirtualProtect|~\cite{virtual-protect}) makes the injected shellcode
    executable. As a result, control is transferred to a regular shellcode.
  \item \textbf{Shellcode stage.} The malicious payload is contained in the shellcode,
    which is now executable. Execution of the shellcode completes the
    exploitation.
\end{enumerate}

Peter Van Eeckhoutte~\cite{vaneeckhoutte10} described a way to bypass DEP in 32-bit Windows programs
with gadget \texttt{pushad ; ret}. Registers are pre-initialized with values so
that a regular ROP chain is on stack after executing the \verb|pushad| instruction (which saves general-purpose registers onto
the stack). In turn, the ROP chain calls the
\verb|VirtualProtect| function to make the stack executable, and transfer control to
the regular shellcode located up the stack. A detailed description of this method and
the Figure can be found in~\cite{vishnyakov18}.

%%%%%%%%%%%%%%%%%%%%%%%%%%%%%%%%%%%%%%%%
\section{BYPASSING DEP AND ASLR WITH ROP}
%%%%%%%%%%%%%%%%%%%%%%%%%%%%%%%%%%%%%%%%

Under certain conditions, it is possible to build a code reuse attack on an application
whose binary code is missing from the attacker. Bittau et al.~\cite{bittau14} give an example of
such attack~-- BROP (blind return-oriented programming). In this paper, the attack model
assumes that the attacked web service handles each incoming request in a separate
process, which is generated by \verb|fork| system call. It means that the memory layout
of the processes handlers does not change. This fact allows the attacker to learn the attacked web service dynamically.  

The authors of this work show that under such conditions it is possible to dynamically
search for gadgets in the memory of the attacked process. The gadget search is carried
out by observing the side effects of the attacked program execution. Instead of the
return address, the test value is placed on stack of the function containing the buffer
overflow. Upon exiting the vulnerable function, control is transferred to this address.
Conceptually at this moment, two fundamental events can
occur: a crash or a pause. A crash happens when any address that is beyond the
executable memory areas is given. A pause occurs due to a delay in program
execution, for instance, after calling \verb|sleep| function. Both events are easily observed
through the connection status with the server. The connection closes or remains open
for a while, respectively. The pause instruction address is essential
for the described attack method since it allows one to find and classify ROP gadgets
dynamically.

The attacker finds:
\begin{enumerate}
  \item $S$~-- the address pausing program execution.
  \item $T$~-- the address knowingly causing the program crash.
\end{enumerate}
For gadget search one takes the trail address $P$. If the chain composed by the
attacker with the trial address leads to a pause with a further crash, then the attacker
assigns the trial gadget to a certain class. Below are examples of such chains:
\begin{itemize}
  \item $P$, $S$, $T$, $T$ ...~-- finds gadgets that do not load the values from the stack such as
    \verb|ret| and~\texttt{xor rax, rax ; ret};
  \item $P$, $T$, $S$, $T$, $T$ ...~-- finds gadgets that load only one word from the stack
    such as \texttt{pop rax ; ret} and~\texttt{pop rdi ; ret};
\end{itemize}

A specific determination of the registers used by each load gadget is performed
according to the side effects of function calls from the linkage table (also detected by a
special procedure~\cite{bittau14}) and system calls (\verb|syscall|). The ultimate goal of building a chain
is to call the \verb|write| system call, which reads the image of the executable file from memory
and sends it to the attacker via network. The attacker analyses the executable file in detail and
constructs the ROP chain that performs the necessary actions.

Snow et al.~\cite{snow13} proposed another example of building a ROP chain for an
application whose binary code is unknown to the attacker~-- JIT-ROP. A distinctive
feature of this work is the conditions of a model attack. The authors believe that the
attacked system contains a set of modern protection tools, such as DEP, ASLR, and even
fine-grained address space randomization at each run of the application~\cite{nurmukhametov18}. However,
the authors also believe that in the application under attack there are many leaks
that reveal the address space of the process. Such attack example is described for
IE browsers. The attack is implemented, for example, through malicious JavaScript code
loaded by the browser together with the web page. Then the attacker can build the ROP
chain just during the attack. All search and classification methods should be
lightweight enough to place them in the script code so that they do
not load the attacked computer critically. With this purpose in mind, the authors adapted
the algorithms proposed by Schwartz et al.~\cite{schwartz11},
replacing the classification method with a heuristic algorithm that works reasonably well
in the presence of a huge amount of binary code in the browser address space.

G\"oktas et al.~\cite{goktas18} proposed an approach to the ROP chain forming that works in the
presence of DEP and randomization of the binary image and all libraries. They called
the basic idea of their approach the ``massage'' of the stack. The key point is that
code pointers are written to the stack during execution (at least return addresses,
and sometimes local variables containing pointers to functions).
The recorded data is not cleared during the return from the function and remains there until further calls that
simply overwrite them with some new values. Moreover, uninitialized local variables
leave values written on stack by previous calls. As a result, with carefully selected
input data, the attacker forms in space, located below the stack of a vulnerable
function, a sequence of pointers to the code, interspersed with a place for
data. Using the vulnerability of writing a single value outside the array one corrects
the low bytes of pointers so that they point to ROP gadgets. If necessary, gadget
parameters are subject to the same changes. When building chains, the authors, by
analogy with the work of Ward et al.~\cite{ward19}, are limited to a memory paragraph with
respect to pointers located on program stack. This forces them also to use gadgets
ending in \verb|call| instructions. The main disadvantage of this method is the complexity and
non-automation of gadget search procedure and creation of such a
program execution path, which would set the values lower on stack so that a
skeleton of the future ROP chain would be formed there.

%%%%%%%%%%%%%%%%%%%%%%%%%%%%%%%%%%%%%%%%
\section{JUMP-ORIENTED PROGRAMMING}
\label{sec:jop}
%%%%%%%%%%%%%%%%%%%%%%%%%%%%%%%%%%%%%%%%

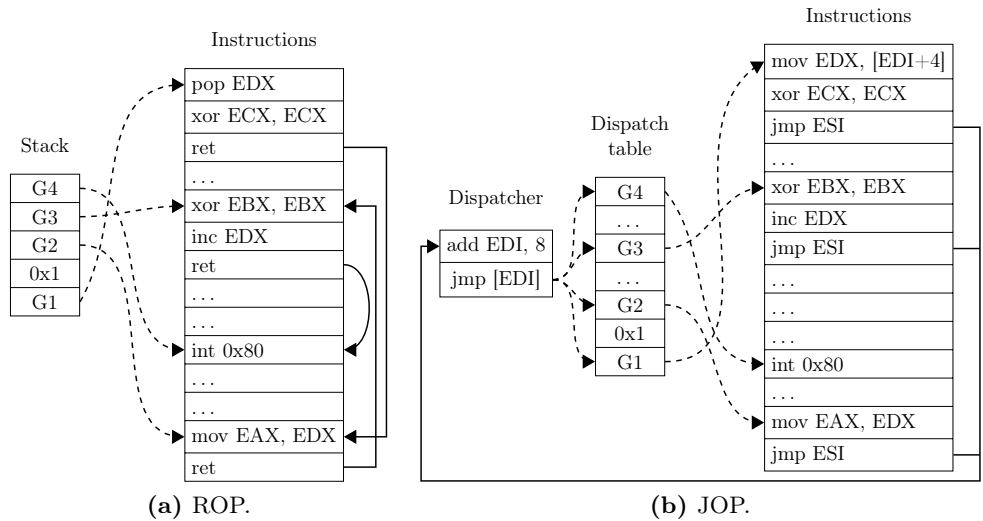
\begin{figure*}[t]
%{{{ rop vs jop
  \centering
  \newcommand\Mstrut{\rule[0em]{0pt}{0.7em}}
  \subfloat[ROP]{
    \centering
    \scalebox{0.7}{
      \begin{tikzpicture}[>=triangle 60]
        \tikzset{stack/.style={draw, anchor=text, rectangle split, rectangle
                                     split parts=5, minimum width=1.3cm }}
        \tikzset{code/.style={draw, anchor=text, rectangle split, rectangle
                                split parts=14, minimum width=1.3cm,
                            rectangle split part align=left}}
        \begin{scope}[every node/.style=stack]
          \node (S) at (0.5,-1){
            \nodepart{one}   \Mstrut G4
            \nodepart{two}   \Mstrut G3
            \nodepart{three} \Mstrut G2
            \nodepart{four}  \Mstrut 0x1
            \nodepart{five}  \Mstrut G1
          };
        \end{scope}
        \node[above=8pt of S] {Stack};
        \begin{scope}[every node/.style=code]
          \node (C) at (3.5,1){
            \nodepart{one}      \Mstrut pop EDX
            \nodepart{two}      \Mstrut xor ECX, ECX
            \nodepart{three}    \Mstrut ret
            \nodepart{four}     \Mstrut \dots
            \nodepart{five}     \Mstrut xor EBX, EBX
            \nodepart{six}      \Mstrut inc EDX
            \nodepart{seven}    \Mstrut ret
            \nodepart{eight}    \Mstrut \dots
            \nodepart{nine}     \Mstrut \dots
            \nodepart{ten}      \Mstrut int 0x80
            \nodepart{eleven}   \Mstrut \dots
            \nodepart{twelve}   \Mstrut \dots
            \nodepart{thirteen} \Mstrut mov EAX, EDX
            \nodepart{fourteen} \Mstrut ret
          };
        \end{scope}
        \node[above=8pt of C] {Instructions};
        % stack -> code
        \draw[->, thick, dashed] (S.five east) to [out=60, in=180] (C.one west);
        \draw[->, thick, dashed] (S.three east) to [out=0, in=180] (C.thirteen west);
        \draw[->, thick, dashed] (S.two east) to [out=0, in=180] (C.five west);
        \draw[->, thick, dashed] (S.one east) to [out=0, in=180] (C.ten west);
        %CFG
        \draw[->, thick] (C.three east) -- ++(0.8,0) |- ++(0,-5) |- (C.thirteen east);
        \draw[->, thick] (C.fourteen east) -- ++(0.6,0) |- ++(0, 4) |- (C.five east);
        \draw[->, thick] (C.seven east) to [out=0, in=0] (C.ten east);
      \end{tikzpicture}
    }
    \label{fig:rop-flow}
  }
  \subfloat[JOP]{
    \centering
    \scalebox{0.7}{
      \begin{tikzpicture}[>=triangle 60]
        \tikzset{table/.style={draw, anchor=text, rectangle split, rectangle
                                     split parts=7, minimum width=1.3cm }}
        \tikzset{code/.style={draw, anchor=text, rectangle split, rectangle
                                split parts=14, minimum width=1.3cm,
                              rectangle split part align=left}}
        \tikzset{dispatcher/.style={draw, anchor=text, rectangle split, rectangle
                                split parts=2, minimum width=1.3cm }}
        \begin{scope}[every node/.style=table]
          \node (T) at (0.6,-1.5){
            \nodepart{one}   \Mstrut G4
            \nodepart{two}   \Mstrut \dots
            \nodepart{three} \Mstrut G3
            \nodepart{four}  \Mstrut \dots
            \nodepart{five}  \Mstrut G2
            \nodepart{six}   \Mstrut 0x1
            \nodepart{seven} \Mstrut G1
          };
        \end{scope}
        \node[above=8pt of T, align={center}] {Dispatch\\table};
        \begin{scope}[every node/.style=code]
          \node (C) at (3.5,1){
            \nodepart{one}      \Mstrut mov EDX, [EDI+4]
            \nodepart{two}      \Mstrut xor ECX, ECX
            \nodepart{three}    \Mstrut jmp ESI
            \nodepart{four}     \Mstrut \dots
            \nodepart{five}     \Mstrut xor EBX, EBX
            \nodepart{six}      \Mstrut inc EDX
            \nodepart{seven}    \Mstrut jmp ESI
            \nodepart{eight}    \Mstrut \dots
            \nodepart{nine}     \Mstrut \dots
            \nodepart{ten}      \Mstrut \dots
            \nodepart{eleven}   \Mstrut int 0x80
            \nodepart{twelve}   \Mstrut \dots
            \nodepart{thirteen} \Mstrut mov EAX, EDX
            \nodepart{fourteen} \Mstrut jmp ESI
          };
        \end{scope}
        \node[above=8pt of C] {Instructions};
        \begin{scope}[every node/.style=dispatcher]
          \node (D) at (-2.6,-2.5) {
            \nodepart{one} \Mstrut add EDI, 8
            \nodepart{two} \Mstrut jmp [EDI]
          };
        \end{scope}
        \node[above=8pt of D] {Dispatcher};
        % table -> code
        \draw[->, thick, dashed] (T.seven east) to [out=0, in=210] (C.one west);
        \draw[->, thick, dashed] (T.five east) to [out=0, in=180] (C.thirteen west);
        \draw[->, thick, dashed] (T.three east) to [out=0, in=180] (C.five west);
        \draw[->, thick, dashed] (T.one east) to [out=310, in=180] (C.eleven west);
        % CFG
        \draw[->, thick] (C.three east) -- ++(0.5,0) |- ++(0,-6.7) |- ++(-10.5, 0) |- (D.one west);
        \draw[-, thick] (C.seven east) -- ++(0.5,0);
        \draw[-, thick] (C.fourteen east) -- ++(0.5,0);
        \draw[->, thick, dashed] (D.two east) to [out=0, in=180] (T.one west);
        \draw[->, thick, dashed] (D.two east) to [out=0, in=180] (T.three west);
        \draw[->, thick, dashed] (D.two east) to [out=0, in=180] (T.five west);
        \draw[->, thick, dashed] (D.two east) to [out=0, in=180] (T.seven west);
      \end{tikzpicture}
    \label{fig:jop-flow}
    }
  }
%}}}
\caption{Comparison of control flow for ROP and JOP with example chain invoking \texttt{exit} system call}
\label{fig:rop-and-jop-flow}
\end{figure*}

Jump-oriented programming~\cite{bletsch11} (JOP) chains gadgets ending
with instructions for jumping to the address controlled by the attacker. In x86 case, it is
such instructions as \texttt{jmp eax} and \texttt{jmp [eax]}. Jump-oriented programming differs by
more complicated procedure of transferring control from one gadget to another
compared to return-oriented programming.

Bletsch et al.~\cite{bletsch11} arrange the control transferring for JOP as follows:
\begin{itemize}
  \item A special gadget type, called a \textit{dispatcher gadget},
    links functional gadgets. This gadget manages
    a virtual program counter and executes a JOP program by moving this counter
    from one functional gadget to another. The functional gadgets addresses are
    stored in the dispatch table.
    Some fixed register is used as a program counter, whose value indicates the cell
    in dispatch table with the current gadget address.
    For example, such gadget is \texttt{add edx, 4 ; jmp [edx]}
    (\verb|edx|~-- virtual program counter).
  \item Gadgets that perform primitive computing operations are called
    \textit{functional}.  Each functional gadget must return control to the dispatcher
    gadget. For example, control return can be implemented by jump through the
    register whose value is always equal to the dispatcher
    gadget address (\texttt{pop eax ; jmp esi}~-- is a gadget that loads \verb|eax|).
\end{itemize}

Figure~\ref{fig:rop-and-jop-flow} schematically shows a control transfer model
in the jump-oriented chain~\ref{fig:jop-flow}
compared with the return-oriented chain~\ref{fig:rop-flow} on the example of a payload that
invokes \texttt{exit(0)} system call.
A JOP chain consists of a set of gadget
addresses and their parameter values stored in memory. It is a dispatch table
with the described control flow. Gadget addresses are essentially opcodes in a virtual
machine defined by the memory state of the attacked process. The ROP chain is
stored on stack, and \verb|esp| register acts as a program counter. JOP differs because
the chain location and the register acting as program counter can be
anything. In the case of JOP in Figure~\ref{fig:jop-flow}, the dispatcher gadget is an instruction
sequence \texttt{add edi, 8 ; jmp [edi]}. The \texttt{edi} register acts as a program counter, which
increases by 8 each time the dispatcher is invoked. The JOP chain is stored in
memory. It is a dispatch table, with the aid of which the dispatcher gadget sequentially
invokes functional gadgets
(G1, G2, G3, G4). Each functional gadget ends in \texttt{jmp ESI}. It allows one to
initially set the value of the \verb|ESI| register to point to the dispatcher gadget.

Formally, the dispatcher gadget can be described as follows:
$$pc \longleftarrow f(pc); goto\ {*}pc; $$
Where a $pc$ can be a register or some address in memory, and $f(pc)$ can be any function
that changes $pc$ in a predictable and monotonous way.

Works~\cite{checkoway10, chen11} used the JOP
trampoline gadget to transfer control from one functional gadget to another (do not
confuse with Section~\ref{sec:stackpivot}). The following instructions can act
as a trampoline gadget: \texttt{pop eax ; jmp [eax]}.
This gadget alternately calls functional gadgets. Functional gadgets can be the following
instruction sequences:
\begin{verbatim}
pop ebx ; jmp [edx]
pop ecx ; jmp [edx]
add ebx, 4 ; jmp [edx]
\end{verbatim}
Each functional gadget performs some basic computing operation and is sure to return
control to the trampoline gadget. To do this, in the given example of instructions, the
address of memory is stored in \verb|edx| register, where the address of the trampoline
gadget is stored. With this approach, the register pressure increases, which may
complicate the construction of chains for x86. However, it may not be a problem
for some other architectures with a large number of registers.

On ARM, the following gadget may act as a trampoline:
\texttt{adds r6, \#4 ; ldr r5, [r6, \#124] ; blx r5}.

Chen et al.~\cite{chen11} showed that from a set of JOP gadgets from real libraries, a Turing-complete
set of instructions can be built.

Apart from jump instructions, at least on the x86 architecture, there are call
instructions \texttt{call eax} and \texttt{call [eax]}. Gadgets that end with such instructions can also be
used at the end of ROP and JOP chains~\cite{bletsch11}.
However, the problem of building a chain of only such gadgets needs a special
approach described in PCOP~\cite{sadeghi17}. Sadeghi et al.~\cite{sadeghi17} show that the direct
application of the method for chain building with a dispatcher gadget, as in
Bletsch et al.~\cite{bletsch11},
is not applicable in this case. When executing a call instruction, two operations
take place:
\begin{enumerate}
  \item The address of the next instruction (return address) is pushed onto the stack.
  \item Control is transferred to the address specified in the instruction.
\end{enumerate}

Using gadgets that end with a call, the main problem is the return addresses placed onto the stack.
They should be removed from the stack by subsequent
gadgets. Sadeghi et al.~\cite{sadeghi17} proposed using \textit{strong trampoline gadgets} for this. Such
gadgets should be located between functional gadgets to transfer control and clear the
stack of useless return address values. In a PCOP chain of $n$ functional gadgets, it is
necessary to place $n-1$ strong trampoline gadgets between functional gadgets. An
example of a strong trampoline gadget can be \texttt{pop x ; pop y ; call y}.
Naturally, Sadeghi et al.~\cite{sadeghi17} demonstrate the Turing completeness of
a set of such gadgets from the \verb|libc| library.

%%%%%%%%%%%%%%%%%%%%%%%%%%%%%%%%%%%%%%%%
\section{SIGRETURN-ORIENTED PROGRAMMING}
%%%%%%%%%%%%%%%%%%%%%%%%%%%%%%%%%%%%%%%%

Bosman et al.~\cite{bosman14} proposed a method for signal handling exploitation in Unix
family operating systems. When a signal is delivered to a process, the kernel
saves its context (registers, stack pointer, processor flags, etc.) in the signal
frame (in user space). As the return address of the signal handler, the kernel places a
pointer to the code that executes the \verb|sigreturn| system call, which
restores the
process context from the signal frame. An attacker can form a signal frame and
call \verb|sigreturn| to change the context of the process. Thus, using only one
gadget that performs the \verb|sigreturn| system call can write arbitrary values to the
registers. This approach is called sigreturn-oriented programming (SROP) and
allows Turing-complete computations. The authors offer two types of attack:
\begin{enumerate}
  \item \textbf{The \texttt{execve} system call.} The attacker places
    \verb|execve| string arguments on stack and forms a signal frame with
    pointers to these strings. Thus, the \verb|sigreturn| gadget initializes
    the registers with arguments of the system call. Moreover, the \verb|syscall|
    gadget, which is already contained in \verb|sigreturn| gadget code,
    invokes the \verb|execve| system call.
  \item \textbf{Fast \texttt{vsyscall}.} Operating systems with a Linux kernel
    version up to 3.3 use the fast system call mechanism known as \verb|vsyscall|. The
    \verb|vsyscall| implementation code contains a set of useful gadgets at fixed
    addresses. In particular, a \verb|syscall| gadget.  The authors claim that gadgets
    remain at the same addresses after kernel security updates and on different
    distributions. Moreover, the same memory page as the \verb|vsyscall| code, keeps
    the current time, whose lower bits, with due patience, can be used as a
    gadget.
\end{enumerate}

%%%%%%%%%%%%%%%%%%%%%%%%%%%%%%%%%%%%%%%%
\section{CONTROL FLOW INTEGRITY}
%%%%%%%%%%%%%%%%%%%%%%%%%%%%%%%%%%%%%%%%

One way of counteracting code reuse attacks is to provide control-flow
integrity (CFI) checking. There is a wide range of publications on this subject that offer
one way or another to implement this method~\cite{burow17}. 
Some of the proposed implementations are even included in the compilers as test
extensions, but they are not used by default for performance reasons.
The general idea of these methods is that during a code reuse attack, the
control flow usually differs significantly from the regular execution control flow.
To detect such deviations, one can build some control flow behavior model
and check its compliance with
actual control transfers during execution. Theoretically, the control flow integrity
prevents the application from being exploited by the code reuse methods described in
previous chapters.

%%%%%%%%%%%%%%%%%%%%%%%%%%%%%%%%%%%%%%%%
\section{USING GADGETS THAT IMMEDIATELY FOLLOW CALL INSTRUCTIONS}
%%%%%%%%%%%%%%%%%%%%%%%%%%%%%%%%%%%%%%%%

\begin{figure}
%{{{ CPROP and LOP figure
  \centering
  \subfloat[CPROP]{
    \centering
    \scalebox{0.7}{
      \begin{tikzpicture}[>=triangle 60]
        \tikzset{side/.style={draw, anchor=text, rectangle split, rectangle
                                     split parts=6, minimum width=1cm }}
        \tikzset{center/.style={draw, anchor=text, rectangle split, rectangle
                                split parts=3, minimum width=1cm }}
        \begin{scope}[every node/.style=side]
          \node (L) at (0,4){
            \nodepart{one} \dots
            \nodepart{two} \dots
            \nodepart{three} call
            \nodepart{four} \dots
            \nodepart{five} \dots
            \nodepart{six} \dots
          };
        \end{scope}
        \begin{scope}[every node/.style=center]
          \node (C) at (2,4){
            \nodepart{one} \dots
            \nodepart{two} \dots
            \nodepart{three} ret
          };
        \end{scope}
        \begin{scope}[every node/.style=side]
          \node (R) at (4,4){
            \nodepart{one} \dots
            \nodepart{two} \dots
            \nodepart{three} call
            \nodepart{four} \dots
            \nodepart{five} \dots
            \nodepart{six} ret
          };
        \end{scope}
        \begin{scope}[every node/.style=side]
          \node (R2) at (6,4){
            \nodepart{one} \dots
            \nodepart{two} \dots
            \nodepart{three} call
            \nodepart{four} \dots
            \nodepart{five} \dots
            \nodepart{six} ret
          };
        \end{scope}
        \draw[->, thick] (L.three east) to [out=0, in=180] (C.one west);
        \draw[->, thick] (C.three east) to [out=0, in=180] (R.four west);
        \draw[->, thick, dashed] (C.three west) to [out=180, in=0] (L.four east);
        \draw[->, thick] (R.six east) to [out=0, in=180] (R2.four west);
      \end{tikzpicture}
      \label{fig:cprop-flow}
    }
  }
  \hspace{1.5cm}
  \subfloat[LOP]{
    \centering
    \scalebox{0.7}{
      \begin{tikzpicture}[>=triangle 60]
        \tikzset{dispatcher/.style={draw, anchor=text, rectangle split, rectangle
                                     split parts=7, minimum width=1cm }}
        \tikzset{gadget/.style={draw, anchor=text, rectangle split, rectangle
                                split parts=3, minimum width=1cm }}
        \begin{scope}[every node/.style=dispatcher]
          \node (D) at (2,4){
            \nodepart{one} \dots
            \nodepart{two} \dots
            \nodepart{three} call
            \nodepart{four} \dots
            \nodepart{five} \dots
            \nodepart{six} jmp
            \nodepart{seven} \dots
          };
        \end{scope}
        \begin{scope}[every node/.style=gadget]
          \node (F1) at (5,5.5){
            \nodepart{one} \dots
            \nodepart{two} \dots
            \nodepart{three} ret
          };
        \end{scope}
        \begin{scope}[every node/.style=gadget]
          \node (F2) at (5,4){
            \nodepart{one} \dots
            \nodepart{two} \dots
            \nodepart{three} ret
          };
        \end{scope}
        \begin{scope}[every node/.style=gadget]
          \node (F3) at (5,2.5){
            \nodepart{one} \dots
            \nodepart{two} \dots
            \nodepart{three} ret
          };
        \end{scope}
        \draw[->, thick, distance=1cm] (D.six west) to [out=180, in=180] (D.one west);
        \draw[->, thick] (D.three east) to [out=0, in=170] (F1.one west);
        \draw[->, thick] (D.three east) to [out=0, in=175] (F2.one west);
        \draw[->, thick] (D.three east) to [out=0, in=190] (F3.one west);
        \draw[->, thick] (F1.three west) to [out=190, in=0] (D.four east);
        \draw[->, thick] (F2.three west) to [out=180, in=0] (D.four east);
        \draw[->, thick] (F3.three west) to [out=170, in=0] (D.four east);
      \end{tikzpicture}
      \label{fig:lop-flow}
    }
  }
  \caption{The control flow of CPROP and LOP. All blocks represent functions}
  \label{fig:cprop-and-lop-flow}
\end{figure}
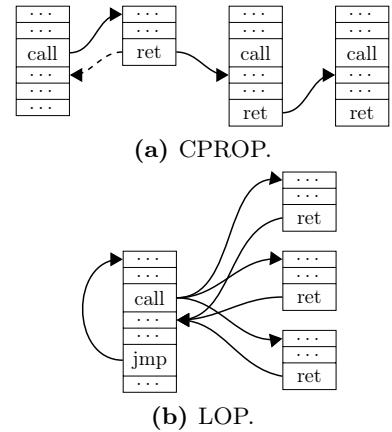

In standard programs, after return from a function, in most cases, control is transferred
to the instruction that goes right after this function call instruction.
In general case, return-oriented programming violates this condition, transferring
control to arbitrary places of the program after the return instruction. To counteract
exploitation by ROP chains some CFI implementations check that return instruction
transfers control to instruction right after the function call instruction. However,
Carlini et al.~\cite{carlini14} noted that in this case, one could use only those gadgets that begin
right after the call instruction (CPROP, Call-Preceded ROP). These gadgets appeared
to be larger, having more complicated side effects. However, the authors showed
that real applications contain these gadgets in quantity sufficient to create workable
ROP chains with a careful account of their side effects. The control flow when using
CPROP gadgets is shown in Fig.~\ref{fig:cprop-flow} by a solid line, and the dashed line shows the original
control flow.

%%%%%%%%%%%%%%%%%%%%%%%%%%%%%%%%%%%%%%%%
\section{REUSE OF FUNCTIONS}
%%%%%%%%%%%%%%%%%%%%%%%%%%%%%%%%%%%%%%%%

\begin{table}[t]
  \caption{A set of POSIX-compliant widgets}
  \small
  \begin{tabular}{l l}
    \toprule
    Category & Widgets \\
    \midrule
    Branching & \makecell[cl]{\texttt{lfind()} + \texttt{longjmp()}, \\
                              \texttt{lsearch()} + \texttt{longjmp()}} \\
    Arithmetic/Logic & \makecell[cl]{\texttt{wordexp()}, \texttt{sigandset()}, \\
                                     \texttt{sigorset()}} \\
    Memory access & \makecell[cl]{\texttt{memcpy()}, \texttt{strcpy()}, \\
                                  \texttt{sprintf()}, \texttt{sscanf()}, etc.} \\
    System calls & \makecell[cl]{\texttt{open()}, \texttt{close()}, \\
                                 \texttt{read()}, \texttt{write()}, etc.} \\
    \bottomrule
  \end{tabular}
  \label{tbl:tc-rilc-widgets}
\end{table}

In order to bypass CFI, specific code reuse attacks are proposed. For example, in
simplest cases, complete functions are enough as gadgets. Tran et al.~\cite{tran11} studied the
question: how expressive the return-to-library attacks are. They show that
return-to-library attacks are, in fact, Turing-complete. They build a
Turing-complete set of \textit{widgets} from POSIX compatible functions of the C standard
library to prove this.
A \textit{widget} is a function with beneficial side effects, so it is an analog of a gadget. To
implement branching, one uses \verb|longjmp()| widgets that change the stack pointer.
Based on a set of widgets, two exploit examples are built. They show the practical
applicability of the proposed approach. Moreover, since this set of widgets is built of
POSIX-compliant functions, widget chains are portable between POSIX-compliant
operating systems. Table~\ref{tbl:tc-rilc-widgets} demonstrates an example of
POSIX-compliant widgets.

It is more challenging to build widget chains than ROP chains manually due to complex
data and control dependencies. Besides, the widget chain requires a larger stack due to the
size of the chain itself.

It is worth noticing that building of exploits consisting of widgets alone is possible on the x86
architecture only with calling convention, in which the function arguments are passed
through the stack. Otherwise, one should use ROP gadgets to load the arguments.

Lan et al.~\cite{lan15} propose a loop-oriented programming (LOP) method that uses whole
functions as gadgets. This method was developed to bypass coarse-grained CFI and the
shadow stack~\cite{davi14}. In order to bypass such security features, it is necessary to transfer control
to the beginning of the function and return control to the calling function at the point
immediately after the call site. The method vaguely resembles jump-oriented
programming (Fig.~\ref{fig:jop-flow}). The authors take whole functions (functional gadgets) as
gadgets, while the dispatcher gadget transfers control between them. A function with a
cycle acts as a dispatcher gadget. Inside the
loop, there is an indirect call. The functional gadgets addresses are stored in the dispatch
table. At each iteration of the loop, the dispatcher gadget monotonously changes the
virtual program counter and calls the next gadget by the address from the dispatch table,
pointed by the counter. After return from the functional gadget, control is
transferred to the next iteration of the dispatcher gadget. Figure~\ref{fig:jop-flow} shows how control is
being transferred between functional gadgets through the dispatcher gadget.

In object-oriented languages, primitives similar to dispatcher functions can often be
met, for example, iterating over collections of same type objects, with a specific virtual
method being called for each of them. For such cases, COOP~\cite{schuster15}, LOOP~\cite{wang18} proposed
a method for exploitation that replaces the virtual call table. Properly created virtual
tables of objects from the collection allow one to organize a chain of objects
methods calls. In this case, by analogy with the previous approaches of the current section,
whole functions act as gadgets. Data can be transferred between different gadgets either
via common fields of objects or through uninitialized variables. In procedural
languages, one may also encounter similar iterations over collections of structures that
contain pointers to processing functions. For an application written in C as an
example, FOP~\cite{guo18} shows an example of building an exploit from program functions.

%}}}

%%%%%%%%%%%%%%%%%%%%%%%%%%%%%%%%%%%%%%%%
\section{DATA FLOW EXPLOITATION}
\label{sec:dop}
%%%%%%%%%%%%%%%%%%%%%%%%%%%%%%%%%%%%%%%%

Chen et al.~\cite{chen05} were the first to show that it is possible to exploit the application data
flow and to execute a payload in its context without breaking the control flow
integrity. The authors presented several real and synthetic examples.
Later, Hu et al.~\cite{hu15, hu16} gave the name to these types of attacks
(DOP, Data-Oriented Programming). They also showed that DOP attacks could be Turing-complete,
i.e., they can execute arbitrary code without violation of the program control
flow integrity. Thus, such attacks can not be detected by control-flow integrity
methods.

When building DOP chains, one uses DOP gadgets that can be arbitrary pieces of code,
and a special dispatcher gadget which is necessary to transfer control between DOP
gadgets. The instructions of the virtual machine in which the DOP chaining is executed
are some instruction sequences in the original program. The values of the DOP chain
variables are stored in memory, since the registers are actively used in the program and
tend to get corrupted between executions of two consecutive DOP gadgets. An example of
a dispatcher gadget is a cycle with some mechanism to select a DOP gadget, which allows
the current gadget to transfer control to the next gadget.

Hu et al.~\cite{hu15, hu16}
semi-automatically created DOP chains. One should find all
DOP gadgets and a dispatcher gadget to build a DOP chain. Having discovered them, one
should find the input data for the target program, which leads the execution path to the
place containing the gadgets found. 
Furthermore, each gadget is supplied with information about which region of memory it has changed
(global variables, function parameters, local variables).
The easiest way is to use gadgets that change the state of global memory. When there are
memory errors in the program, attack designing is divided into the following stages:
searching for gadgets, selecting suitable gadgets, and stitching them. To stitch gadgets, Hu et
al.~\cite{hu15} build a 2D data flow graph that represents data flows in two dimensions: memory
addresses and runtime. Then they try to discover new edges on this graph. The authors
show some examples of designed attacks on real applications that bypass DEP, ASLR, and
CFI protections. Besides, they show that some real applications contain enough
gadgets to create DOP chains, including Turing-complete ones.

In works~\cite{pewny19, ispoglou18}, DOP and methods for the automated generation of DOP chains got
further development. Despite the similarity of the main idea of this type of attacks, the
methods for the automated generation of such chains are significantly different. Their
detailed discussion is beyond the scope of this article.

%%%%%%%%%%%%%%%%%%%%%%%%%%%%%%%%%%%%%%%%
\section{GENERAL CODE REUSE EXPLOIT GENERATION SCHEME}
\label{sec:generation-scheme}
%%%%%%%%%%%%%%%%%%%%%%%%%%%%%%%%%%%%%%%%

We schematically divide the process of code reuse exploits generation into four stages:
\begin{enumerate}
  \item The gadget search in non-randomized executable areas of the process memory
    image (Section~\ref{sec:gadget-search}).
  \item The gadget semantics determination (some methods may skip this stage).
    At this stage, the payload that each gadget performs is determined (Section~\ref{sec:semantic-definition}).
  \item The combination of gadgets and their parameters to obtain a chain of
    gadgets that performs a given sequence of actions (Section~\ref{sec:chain-generation}).
  \item The automated exploit generation~\cite{avgerinos11, cha12,
    padaryan15, fedotov16, shoshitaishvili16}~-- generation of input data causing the
    program exploitation by injecting and executing the ROP chain. At this
    stage, machine instructions (on the program execution trace from the point
    of receiving input data to the vulnerable point) are symbolically
    executed~\cite{king76, schwartz10, godefroid08}. Thus, a path predicate is
    constructed. The path predicate is united with the
    security predicate that describes the ROP chain injection and
    transferring the control to it. The solution to the obtained system of
    equations is an exploit. The path predicate ensures that the program runs
    the same path to vulnerability, and the security predicate provides control
    flow hijacking.
\end{enumerate}

%%%%%%%%%%%%%%%%%%%%%%%%%%%%%%%%%%%%%%%%
\section{GADGET CATALOG}
\label{sec:catalogue}
%%%%%%%%%%%%%%%%%%%%%%%%%%%%%%%%%%%%%%%%

\begin{table*}[t]
  \caption{Incomplete gadget catalog}
  \small
  \begin{tabular}{l l l l l}
    \toprule
    Semantic              & Virtual      & Machine     & Gadget    & Side   \\
        description       &  address     & description & parameters & effects  \\
    \midrule
    r1 += r2              &              &             &           & \\
    \midrule
    r = M[ESP + Offset]   &              &             &           & \\
    \midrule
    r1 = M[ESP + Off1]    &              &             &           & \\
    r2 = M[ESP + Off2]    &              &             &           & \\
    r3 = M[ESP + Off3]    &              &             &           & \\
    \bottomrule
  \end{tabular}
  \label{tbl:gadget_catalogue_example_1}
\end{table*}

\begin{table*}[t]
  \caption{Complete gadget catalog}
  \small
  \begin{tabular}{l l l l l}
    \toprule
    Semantic              & Virtual      & Machine     & Gadget    & Side   \\
        description       &  address     & description & parameters & effects  \\
    \midrule
    r1 += r2              & 0xdeadbeef & add eax, ebx  & r1 = eax  & edx \xmark\\
                          &            & pop edx       & r2 = ebx  &           \\
                          &            & ret           &           &           \\
                          & 0xcafecafe & add eax, ebx  & r1 = eax  & ecx \xmark\\
                          &            & pop ecx       & r2 = ebx  &           \\
                          &            & ret           &           &           \\
                          & 0xcafebabe & add edx, ecx  & r1 = edx  & ~---      \\
                          &            & ret           & r2 = ecx  &           \\
    \midrule
    r = M[ESP + Offset]   & 0x12345678 & pop eax       & r = eax   & ~---      \\
                          &            & ret           & Offset = 0&           \\
                          & & & & \\
    \midrule
    r1 = M[ESP + Off1] & 0x10203040 & pop eax          & r1 = eax, Off1 = 0 & ~--- \\
    r2 = M[ESP + Off2] &            & pop ebx          & r2 = ebx, Off2 = 4 &      \\
    r3 = M[ESP + Off3] &            & pop ecx          & r3 = ecx, Off3 = 8 &      \\
                       &            & ret              &                    &      \\
    \bottomrule
  \end{tabular}
  \label{tbl:gadget_catalogue_example}
\end{table*}

Before discussing specific methods for searching and determining the
gadgets semantics, let us define the \textit{gadget catalog} as a list of entries with the following
contents:
\begin{enumerate}
  \item \textbf{Semantic description} of the instruction sequence. Each
    description usually corresponds to some basic computational or memory
    operation (addition, subtraction, writing to memory, reading from memory,
    initializing the register with some value, transferring control, etc.).
  \item \textbf{Virtual address} of the gadget found in the application address
    space. It is an operation code for the instruction set architecture that is
    defined by the gadget catalog.
  \item \textbf{Machine instructions} of the gadget are a specific instruction
    sequence that implements a given semantic description. They can be cataloged
    manually or automatically during application binary image analysis.
  \item \textbf{Gadget parameters} are parameters of the semantic description
    (specific registers, constants, etc.).
  \item \textbf{Side effects} of gadget execution relatively to its semantics.
    A side effect is any change in memory and registers, not described by the
    gadget semantics. Side effects can be cataloged manually or automatically
    calculated during gadgets classification.
\end{enumerate}

We give an example to clarify this definition. Table~\ref{tbl:gadget_catalogue_example_1} presents the gadgets
catalog, consisting of several semantic descriptions. The first semantic description
corresponds to the operation of adding the values of two registers $r1 \mathbin{+=} r2$. The second
semantic description corresponds to the instruction for loading a value from the stack
into the register. The last semantic description defines the gadget loading three
registers from the stack. After the gadget search, the catalog takes the form shown in
Table~\ref{tbl:gadget_catalogue_example}. The first two gadgets, found at addresses 0xdeadbeef and 0xcafecafe, have
side effects relative to the basic semantic description because they change
values of \verb|edx| and \verb|ecx| registers respectively.

%%%%%%%%%%%%%%%%%%%%%%%%%%%%%%%%%%%%%%%%
\subsection{Turing-Complete Gadget Catalog}
%%%%%%%%%%%%%%%%%%%%%%%%%%%%%%%%%%%%%%%%

The authors of several works~\cite{homescu12, chen11, sadeghi17, shacham07, roemer12, tran11}
compose a gadget catalog in such
a way that the set of semantic descriptions is Turing-complete. Such gadget catalog
defines some new computing machine capable of performing arbitrary computations.

While searching and determining gadgets semantics, their addresses are being
cataloged. After that, two situations are possible:
\begin{enumerate}
  \item For each semantic description, a specific gadget has been found.
  \item There are no specific gadgets for some semantic descriptions.
\end{enumerate}

In the first case, it turns out that the found set of addresses implements the computer
described by the catalog on a specific executable file. It means that it is possible to
make arbitrary calculations with gadgets found. Moreover, one can use this catalog
to describe the target architecture instruction set for the C compiler (llvm~\cite{roemer12}).

In the
second case, when there are no gadgets for some semantic descriptions in the Turing-complete
gadget catalog, arbitrary calculations can no longer be
performed. Therefore, the question arises: what is the computational power of found
gadgets set? In other words, is it possible to execute a given program with the found set
of gadgets? To answer this question, each program that describes the exploit should be
analyzed separately. Based on the operations that it performs and
the gadget catalog contents, in some cases, it can be concluded before the exploit
generation that generation is impossible. For example, there are conditional transitions
in the initial exploit program, while there are no gadgets that
implement branching in the gadget catalog. A similar situation is with writing to memory. In other cases,
one needs to try to build an exploit from existing gadgets. If it is built, then the answer
to the question is positive and is confirmed by the designed exploit. Otherwise, the
generation task can be reduced to an exhaustive search of all possible combinations of
gadgets, which can be time-consuming for a large gadget catalog. We suppose that the
infinite exhaustive search with no possibility of generating an exploit is rare for real
executable files.

In order to compile Turing-complete ROP chains, one should be able to
conditionally change the stack pointer that acts as a program counter. Roemer et al.~\cite{roemer12}
propose the following way to implement conditional branching for the x86
architecture:
\begin{enumerate}
  \item Perform some operation that updates the interesting flag.
  \item Copy the interested flag from the flag register to the general-purpose register.
  \item Use this flag for conditional changing the stack pointer by the desired
    offset (for instance, by multiplying the offset by the flag value 0 or 1).
\end{enumerate}

%%%%%%%%%%%%%%%%%%%%%%%%%%%%%%%%%%%%%%%%
\section{GADGET SEARCH}
\label{sec:gadget-search}
%%%%%%%%%%%%%%%%%%%%%%%%%%%%%%%%%%%%%%%%

Regardless of the exploit building process, one should first find all available gadgets in the
binary image of the application. There are two fundamental approaches to the gadget
search task. The first of them offers to search for gadgets by the list of templates. Templates
are usually specified by regular expressions over binary codes of gadget commands.
Initially, the gadget catalog contains semantic descriptions of gadgets. For each
semantic description, gadgets are searched according to some template. As a result,
specific gadgets are included in the gadget catalog for corresponding semantic descriptions: virtual
addresses, machine instructions, and gadget parameters. Side effects (for instance,
clobbered registers~\cite{monapy, ropper}) can be obtained upon analyzing the machine instructions
of found gadgets.

The second approach is automatic search for all possible instruction sequences
ending in a control transfer instruction. The classic algorithm that implements the
search for all gadgets is Galileo algorithm~\cite{shacham07}. It first looks for control
transfer instructions in the executable sections of the program. For each instruction
found, it tries to disassemble several bytes preceding the instruction. All correctly
disassembled instruction sequences are cataloged. Thus, the catalog
contains virtual addresses and machine instructions for gadgets.
Many open-source gadget search tools use this algorithm~\cite{ropgadget, monapy, barf,
angrop, roper, pshape, ropper, ropgenerator,
mipsidaplugin, roptools, idarop, exrop}.

%%%%%%%%%%%%%%%%%%%%%%%%%%%%%%%%%%%%%%%%
\section{DETERMINING GADGET SEMANTICS}
\label{sec:semantic-definition}
%%%%%%%%%%%%%%%%%%%%%%%%%%%%%%%%%%%%%%%%

Not all found gadgets are suitable for building ROP chains. One should
understand what useful payload this gadget performs in order to use the gadget
in building the ROP chain. The gadget semantics
can be determined manually~\cite{shacham07}. In the template
search for gadgets, the semantics are contained in the template
description~\cite{hund09, homescu12, goktas18, buchanan08, sadeghi17, chen11,
bletsch11, checkoway10, roemer12}.

%%%%%%%%%%%%%%%%%%%%%%%%%%%%%%%%%%%%%%%%
\subsection{Gadget Types}
\label{sec:gadget-types}
%%%%%%%%%%%%%%%%%%%%%%%%%%%%%%%%%%%%%%%%

Schwartz et al.~\cite{schwartz11} proposed defining the functionality of the gadget by its
belonging to some parameterized types that define the new instruction set
architecture (ISA). Type parameters are registers, constants, and binary
operations. In order to use the gadget when building ROP chains, one should
fulfil the following gadget properties:
\begin{itemize}
  \item \textbf{Functional.} Each gadget has a type that determines its
    functionality. The type of gadget is described semantically with the
    postcondition~-- the Boolean predicate $\mathcal{B}$. It must always be
    true after the gadget is executed. It is worth noticing that one gadget can
    belong to several types simultaneously. For example, the gadget
    \texttt{push eax ; pop ebx ; pop ecx ; ret} simultaneously moves \verb|eax|
    into \verb|ebx| and loads the value from the stack into \verb|ecx|, which
    corresponds to \texttt{MoveRegG: ebx $\leftarrow$ eax} and
    \texttt{LoadConstG: ecx $\leftarrow$ [esp + 0]}.
  \item \textbf{Control preserving.} Each gadget should be able to transfer
    control to another gadget.
  \item \textbf{Known side effects.} The gadget should not have unknown side
    effects. Side effects of gadget execution should not lead to uncontrolled
    program behavior. For example, writing a value to an arbitrary memory
    address may cause a program crash.
  \item \textbf{Constant stack offset.} Most gadget types require the stack
    pointer to increase by a constant value after each execution.
\end{itemize}

In order to determine whether the gadget instruction sequence $\mathcal{I}$ satisfies the
postcondition $\mathcal{B}$, Schwartz et al.~\cite{schwartz11} use a well-known technique from formal
verification, namely, computing the weakest precondition~\cite{jager10}. At a
high level, the
weakest precondition $wp(\mathcal{I},\mathcal{B})$ for an instruction sequence
$\mathcal{I}$ and condition $\mathcal{B}$ is a
boolean precondition that describes when $\mathcal{I}$ is completed in the state satisfying $\mathcal{B}$.

The weakest precondition is used to ensure that the gadget semantics are always
determined after the execution of the instruction sequence $\mathcal{I}$. For this purpose, it is
sufficient to check:
\begin{equation*}
  wp(\mathcal{I}, \mathcal{B}) \equiv true
\end{equation*}
If the formula is correct, then $\mathcal{B}$ is always true after performing $\mathcal{I}$,
which means that $\mathcal{I}$ is a gadget with semantic type $\mathcal{B}$.

However, formal verification of the gadget semantics practically appeared to be very
slow. The authors proposed a combined approach to speedup the process. The gadget
instructions are preliminarily executed several times on random input data, and then
the truth of $\mathcal{B}$ is verified. If B turns out to be false for at least one execution, then the
instruction sequence cannot be a gadget of this type. Thus, a more complicated
computation of the weakest precondition is performed only if $\mathcal{B}$ is true for each
execution.

The combined approach can be conventionally divided into two stages: gadget
classification and gadget verification. At the classification stage, hypotheses are made
that gadgets belong to certain types and about the values of these types parameters.
Hypotheses are essentially defined by Boolean postconditions. And at the verification
stage, for each postcondition, its truth or falsehood is formally proved, and
the hypothesis is accepted or rejected, respectively.

%%%%%%%%%%%%%%%%%%%%%%%%%%%%%%%%%%%%%%%%
\subsubsection{Gadget Classification}
%%%%%%%%%%%%%%%%%%%%%%%%%%%%%%%%%%%%%%%%

Currently, there are a lot of processor architectures with various instructions.
In order to abstract from the specifics of a particular architecture when writing
universal algorithms, one traditionally uses an intermediate representation of machine
instructions (VEX~\cite{nethercote07}, REIL~\cite{dullien09}, Pivot~\cite{padaryan10, soloviev19}, etc.).
In this case, the binary code
analysis algorithms work with a simpler intermediate representation, and not with the
target processor architecture.

In papers~\cite{heitman14, vishnyakov18}, gadget classification is based on the interpretation of gadget instructions
intermediate representation. During interpretation, accesses to registers and memory
are tracked. If the first reading of a register or memory area occurs, the value read is
randomly generated. As a result of the interpretation, the initial and final values of the
registers and memory are obtained. Based on this information, possible
gadget types are guessed.
For example, for
membership in the \verb|MoveRegG| type~\cite{schwartz11}, such pair of registers should exist, that the
initial value of the first register equals to the final value of the second one. The analysis
results in a list of all types satisfying the gadget, and their parameters (list of candidates).
Then, several more runs of the interpretation process with diverse input data are
performed. Thus, erroneously determined types are removed from the list
of candidates.

Moreover, the gadget classification may result in the following~\cite{vishnyakov18}:
\begin{itemize}
  \item The list of ``clobbered'' registers, whose values changed during the gadget execution.
  \item Information about a gadget frame (Section~\ref{sec:gadget-frame}): the frame size and the
    offset of the cell with the address of the next gadget relative to the
    beginning of the frame.
\end{itemize}

It is worth noticing that the number of incorrectly classified gadgets can be reduced by
adding runs of the interpretation process with boundary input data 0 and -1. The
percent of incorrectly classified gadgets, in this case, is insignificant and amounts to
0.7\%~\cite{vishnyakov19}.

%%%%%%%%%%%%%%%%%%%%%%%%%%%%%%%%%%%%%%%%
\subsubsection{Gadget Verification}
%%%%%%%%%%%%%%%%%%%%%%%%%%%%%%%%%%%%%%%%

\begin{table*}[t]
  \caption{Verification of gadget \texttt{ArithmeticLoadG: ebx $\leftarrow$ ebx + [eax]}}
  \centering
  \footnotesize
  \begin{tabular}{l l l l}
    \toprule
    \textbf{Step} & \textbf{Symbolic state} & \textbf{Instruction} &
    \textbf{Set of formulas} \\
    \midrule
    initial &
    \makecell[cl]{$M$, $eax = \phi_1$, $ebx = \phi_2$, \\
                  $ecx = \phi_3$, $esp = \phi_4$, \\
                  $eip = \phi_5$} & --- & $S_0 = \emptyset$ \\
    1 & $ecx = \phi_6$ & \texttt{mov ecx, [eax]} &
    \makecell[cl]{$S_1 = S_0 \cup \{\phi_6 = (concat$ \\
      ~~~~$(select\ M\ \phi_1)$ \\
      ~~~~$(select\ M\ \phi_1 + 1)$ \\
      ~~~~$(select\ M\ \phi_1 + 2)$ \\
      ~~~~$(select\ M\ \phi_1 + 3))\}$} \\
    2 & $ebx = \phi_7$ & \texttt{add ebx, ecx} &
    $S_2 = S_1 \cup \{\phi_7 = \phi_2 + \phi_6\}$ \\
    final & $eip = \phi_8$, $esp = \phi_9$ & \verb|ret| &
    \makecell[cl]{$S_3 = S_2 \cup \{\phi_8 = (concat$ \\
      ~~~~$(select\ M\ \phi_4),$ \\
      ~~~~$(select\ M\ \phi_4 + 1),$ \\
      ~~~~$(select\ M\ \phi_4 + 2),$ \\
      ~~~~$(select\ M\ \phi_4 + 3)),$ \\
      ~~~~$\phi_9 = \phi_4 + 4\}$} \\
    \midrule
    & \multicolumn{2}{l}{\textbf{Determining semantics}} &
    \textbf{Verification} \\
    \midrule
    verify & \multicolumn{2}{l}{
      $final(ebx) = initial(ebx) + initial(M[eax])$} &
      \makecell[cl]{$\phi_7 \neq \phi_2 + (concat$ \\
                    ~~~~$(select\ M\ \phi_1)$ \\
                    ~~~~$(select\ M\ \phi_1 + 1)$ \\
                    ~~~~$(select\ M\ \phi_1 + 2)$ \\
                    ~~~~$(select\ M\ \phi_1 + 3))$ \\
                    is \textbf{UNSAT}} \\
    \bottomrule
  \end{tabular}
  \label{tbl:verification}
\end{table*}

\begin{listing}[t]
\begin{lstlisting}[
    mathescape,
    basicstyle=\ttfamily\small,
    columns=fullflexible,
    keepspaces=true
]
neg eax ; sbb eax, eax ; and eax, ecx ;
pop ebp ; ret
MoveRegG: EAX $\leftarrow$ ECX
\end{lstlisting}
\caption{Incorrectly classified gadget that is rejected after verification.}
\label{lst:verifier}
\end{listing}

Gadget classification is a set of postconditions that describes the possible
gadget semantics. Gadget verification allows one to formally prove the truth of these
postconditions for arbitrary input data. The gadget can be verified by
the weakest precondition~\cite{schwartz11, ouyang15} or through a
symbolic execution of the gadget instructions~\cite{heitman14, vishnyakov18-slides, vishnyakov19}.

We consider the method for gadget verification through symbolic
execution~\cite{king76, schwartz10, godefroid08} in detail.
During symbolic execution, the gadget semantics are modelled with SMT~\cite{smt-lib}
expressions. Initially, all registers are assigned to free symbolic variables.
The symbolic
memory at the beginning is an empty byte array $M$ of bit vectors:
$$M = (Array\ (\_\ BitVec\ \langle addrSize \rangle) \ (\_\ BitVec\ 8)),$$
where $\langle addrSize \rangle$~-- the dimension of the architecture address word. The symbolic
state contains a mapping from registers to symbolic variables and the current state of
symbolic memory. Symbolic execution of gadget instructions generates SMT formulas
over variables and constants, and also updates the symbolic state of registers and
memory in accordance with the instruction operational semantics. Operation
with symbolic memory is performed through $select$ and $store$ operations on $Array$.
The function $(select\ M\ i)$ returns the $i$-th element of the array $M$ and simulates reading a byte at
address $i$. The function $(store\ M\ i\ b)$ returns the array obtained from the array $M$ by
storing element $b$ at index $i$, which simulates a record of byte $b$ at address $i$.

Heitman et al.~\cite{heitman14} first translate the gadget instructions into an intermediate REIL
representation~\cite{dullien09}. And only after that, REIL instructions are subjected to symbolic
execution.

Verification postcondition is a Boolean predicate over initial and final values
of registers and memory. Registers and memory from the corresponding symbolic states are substituted
in the predicate. The validity of the postcondition formula is verified through the
unsatisfiability of its negation with the SMT solver.

Table~\ref{tbl:verification} shows an example of
\texttt{ArithmeticLoadG: ebx $\leftarrow$ ebx + [eax]} verification.
Initially, registers are assigned to free symbolic variables $\phi_i$, and
an array represents the memory. A set of formulas is empty. New formulas are added in
accordance with the operational semantics of the instruction under interpretation.
Formulas are created according to SSA form~-- when a formula is added, a new symbolic
variable is created to which this formula is assigned. At the first step, a new symbolic
variable $\phi_6$ is created. It is equal to the value of the second instruction operand
\verb|[eax]| loaded from memory. In the symbolic state, the \verb|ecx| register is assigned the symbolic
variable $\phi_6$. At the second step, the result of addition is assigned to the variable
$\phi_7 = \phi_2 + \phi_6$, which, in turn, is assigned to the resultant
instruction operand~-- the $ebx$
register in the symbolic state. At the final step, the symbolic state is updated according to
the return instruction operational semantics, i.e., the instruction pointer is loaded
from the stack, and the stack pointer is incremented by 4. Symbolic variables from the
initial and final symbolic states are substituted in the postcondition describing the gadget
type. SMT solver checks the satisfiability of the formula negation.
Negation of the formula is unsatisfiable, which means the gadget satisfies the declared
type with parameters.

Listing~\ref{lst:verifier} shows an example of a gadget that can be misclassified,
and verification fixes this error. During classification, the gadget was classified as
\texttt{MoveRegG: EAX $\leftarrow$ ECX}. For the nonzero initial value of the \verb|eax| register, the gadget
copies the value of the \verb|ecx| register to \verb|eax|. However, if the initial value of \verb|eax| is
zero, then its final value is zero, which is not a copy of \verb|ecx|
register value
other than zero.

%%%%%%%%%%%%%%%%%%%%%%%%%%%%%%%%%%%%%%%%
\subsubsection{Gadget Cataloging}
%%%%%%%%%%%%%%%%%%%%%%%%%%%%%%%%%%%%%%%%

Gadgets are cataloged (Section~\ref{sec:catalogue}) as follows. The gadget catalog initially
contains semantic descriptions of gadget types. Each gadget found by the Galileo
algorithm is classified. As a result of the classification, one obtains semantic
descriptions (gadget types) that correspond to the gadget. Entries with gadget virtual
addresses and machine instructions are appended to the corresponding
semantic descriptions.
Also, gadget parameters (values of type parameters) and side effects (``clobbered'' registers)
are cataloged in these entries. Further, all gadgets
from the catalog are verified. As a result of verification, incorrectly classified gadgets
are deleted from the catalog.

%%%%%%%%%%%%%%%%%%%%%%%%%%%%%%%%%%%%%%%%
\subsection{Gadget Summary}
\label{sec:summary}
%%%%%%%%%%%%%%%%%%%%%%%%%%%%%%%%%%%%%%%%

\textit{A gadget summary}~\cite{follner16, kornau10, dullien10} is a description of gadget semantics in the form of a
compact specification.  Gadget summaries contain preconditions and postconditions over
the values of registers and memory. In particular, the gadget summary may contain:
\begin{itemize}
  \item registers loaded from the stack (\texttt{eax = [esp + 4]}),
  \item registers read from memory (\texttt{ecx = [edx + 2]}),
  \item registers whose values were changed (\texttt{ecx = eax + ebx}),
  \item ranges of memory addresses used for reading or writing
        (\texttt{[rsp] <-> [rsp + 0x20]}).
\end{itemize}

Follner et al.~\cite{follner16} proposed the following method for composing a gadget summary.
First, gadget instructions rise to the level of VEX~\cite{nethercote07} intermediate representation.
Then all assignments are advanced to form a single expression, called a
postcondition. The postcondition describes all operations by which the final
value was obtained in the register under consideration. The analysis supports a memory
model that allows one to simulate the situation of passing values via the stack correctly.
Also, this analysis allows one to obtain preconditions that describe ranges of memory
accesses by register with an offset (\texttt{[rax] <-> [rax + 0x20]}). Preconditions
indicate that registers from these memory ranges should point to memory available for
reading or writing.

\textbf{Gadgets cataloging} (Section~\ref{sec:catalogue}) goes as follows. The
gadget catalog initially contains gadgets virtual addresses and machine instructions. A
summary is composed for each gadget from the catalog, which essentially allows
one to catalog
gadget semantic description and side effects.

%%%%%%%%%%%%%%%%%%%%%%%%%%%%%%%%%%%%%%%%
\subsection{Gadget Dependency Graph}
\label{sec:gadget-graph}
%%%%%%%%%%%%%%%%%%%%%%%%%%%%%%%%%%%%%%%%

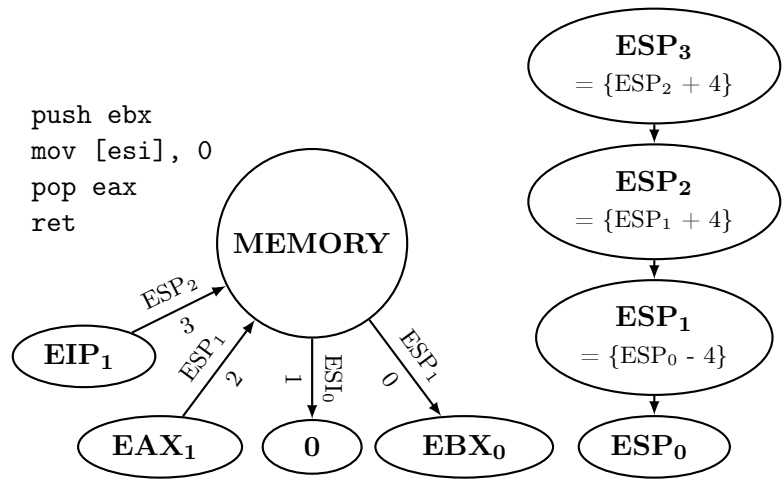
\begin{figure*}[t]
  %{{{
    \centering
\tikzset{
    >=latex,
    n/.style = {shape=ellipse,draw,thick,align=center,minimum width=1.3cm},
    c/.style = {shape=circle,draw,thick,align=center,minimum width=2.5cm},
    g/.style = {align=left}
}
    \begin{tikzpicture}
      [
        every edge/.style = {draw, ->, thick},
        sloped
      ]
      \node[c](mem) at (0,0.5) {\textbf{MEMORY}};
      \node[n](eax1) at (-2,-2.2) { \textbf{EAX\textsubscript{1}} };
      \node[n](eip1) at (-3,-1) { \textbf{EIP\textsubscript{1}} };
      \node[n](zero) at (0,-2.2) {\textbf{0}};
      \node[n](ebx0) at (2,-2.2) {\textbf{ EBX\textsubscript{0}} };
      \draw (eax1) edge node[midway, above] {\footnotesize ESP\textsubscript{1}}
        node[midway, below] {\footnotesize 2} (mem);
      \draw (eip1) edge node[midway, above] {\footnotesize ESP\textsubscript{2}}
        node[midway, below] {\footnotesize 3} (mem);
      \path (mem) edge node[midway, above] {\footnotesize ESI\textsubscript{0}}
        node[midway, below] {\footnotesize 1} (zero);
      \path (mem) edge node[midway, above] {\footnotesize ESP\textsubscript{1}}
        node[midway, below] {\footnotesize 0} (ebx0);
      \node[n](esp3) at (4.5, 2.85) {\textbf{ESP\textsubscript{3}}\\
        \footnotesize= \{ESP\textsubscript{2} + 4\}};
      \node[n](esp2) at (4.5, 1.05) {\textbf{ESP\textsubscript{2}}\\
        \footnotesize= \{ESP\textsubscript{1} + 4\}};
      \node[n](esp1) at (4.5, -0.75) {\textbf{ESP\textsubscript{1}}\\
        \footnotesize= \{ESP\textsubscript{0} - 4\}};
      \node[n] (esp0) at (4.5, -2.2) { \textbf{ESP\textsubscript{0}} };
      \draw (esp3) edge (esp2);
      \draw (esp2) edge (esp1);
      \draw (esp1) edge (esp0);
      \node[g](gad) at (-2.5,1.5) {\texttt{push ebx} \\
                                   \texttt{mov [esi], 0} \\
                                   \texttt{pop eax} \\
                                   \texttt{ret}};
    \end{tikzpicture}
    \caption{Gadget dependency graph}
    \label{fig:gadget-graph}
    %}}}
\end{figure*}

Milanov~\cite{milanov18} proposed to represent a gadget in the form of a directed dependency
graph (Fig.~\ref{fig:gadget-graph}). Vertices correspond to registers, memory, and constants. All memory is
represented by one node. While a register can correspond to several vertices: each
modification of the register generates a new vertex ($reg_0$, $reg_1$, $reg_2$, etc.). Directed edges
reflect data dependencies (register assignment, memory access, etc.). Gadget
instructions generate new edges on the graph. The edges connected to the memory
also contain tags with a memory access address and are numbered in chronological
order.

\textbf{Gadget cataloging} (Sec.~\ref{sec:catalogue}) goes as follows. Initially, the catalog contains virtual
addresses and gadget instructions. The instructions of each gadget are translated into
REIL~\cite{dullien09} intermediate representation, for which a dependency graph is constructed.
As a result of the graph traversal, a gadget semantic description is computed: the
final values of registers and memory are expressed via the initial ones. An
expression for some finite value may have a condition under which this expression is
true. Further, the gadgets are classified by type (Section~\ref{sec:gadget-types}). The author’s motivation
for such method of determining the gadget semantics was a shorter working time
compared to methods using SMT-solvers.

For example in Figure~\ref{fig:gadget-graph} the following semantic description
is obtained:

\begin{small}
\begin{align*}
  EIP_1 &= MEM[ESP_0] \\
  ESP_3 &= ESP_0 + 4 \\
  EAX_1 &=
  \begin{cases}
    0 & \quad \text{if } ESP_0 - 4 = ESI_0 \\
    EBX_0 & \quad \text{if } ESP_0 - 4 \neq ESI_0
  \end{cases}
  \\
  MEM[ESP_0 - 4] &=
  \begin{cases}
    0 & \quad \text{if } ESP_0 - 4 = ESI_0 \\
    EBX_0 & \quad \text{if } ESP_0 - 4 \neq ESI_0
  \end{cases}
  \\
  MEM[ESI_0] &= 0
\end{align*}
\end{small}

%%%%%%%%%%%%%%%%%%%%%%%%%%%%%%%%%%%%%%%%
\section{GADGET CHAINS GENERATION}
\label{sec:chain-generation}
%%%%%%%%%%%%%%%%%%%%%%%%%%%%%%%%%%%%%%%%

This section describes various methods for ROP chains generation.
It is worth noticing that chaining gadgets is a brute force task, therefore, to reduce the
number of brute-force iterations, one can prefilter unnecessary gadgets and sort
them by quality~\cite{follner16_2}. ROP chains generation differs from regular compilation in the
following ways:
\begin{itemize}
  \item Most often, a ROP chain cannot save registers values
    in memory for their further recovery due to the lack of
    relevant gadgets.
  \item ROP gadgets may have side effects. For example, a gadget can ``clobber''
    registers. Values of ``clobbered'' registers are not saved after the
    gadget execution. Side effects should be considered during gadgets scheduling~\cite{schwartz11}.
  \item Some gadget types (Section~\ref{sec:gadget-types}) that act as virtual machine
    instructions may not be available. In this case, it is necessary to replace
    the missing gadgets with a sequence of others~\cite{schwartz11}.
\end{itemize}

During generation, one should consider restricted symbols that cannot be used in the
ROP chain. For instance, an overflow may take place with \verb|strcpy|
function, which prevents the chain from containing zero bytes. However, only a few
completely solved the problem of restricted symbols~\cite{ding14}. Most solutions just delete
gadgets whose addresses contain restricted symbols but do not check the gadget
parameter values on stack.

ROP payload can be divided into the following: setting registers to specified
values and execution of one more gadget~\cite{angrop}. Thus, the method for ROP
chains generation can be based on registers setting, and the rest of the payload can be
implemented by appending to the resulting chain one gadget that writes to memory,
calls a function, invokes a system call, etc.

%%%%%%%%%%%%%%%%%%%%%%%%%%%%%%%%%%%%%%%%
\subsection{Turing-Complete Compilation with a Fixed Gadget Catalog}
%%%%%%%%%%%%%%%%%%%%%%%%%%%%%%%%%%%%%%%%

Let us consider building a compiler based on a fixed gadget catalog. Buchanan et al.~\cite{buchanan08, roemer12}
manually composed a Turing-complete gadget catalog from the
machine code of Solaris OS \verb|libc| standard library for the SPARC
architecture. Each semantic description is associated with a single instruction
sequence from the library machine code. SPARC allows only
aligned accesses to
instructions, so all gadgets are legitimate epilogues of library functions.

SPARC
uses register windows. The register window consists of
registers intended for input parameters, return values, and temporary values inside the
procedure. When the function is called, the register window is shifted forward,
and it is reversed on function return. When call stack is large, there is a
lack of register windows, which makes it necessary to save them on stack. In this case,
during function return, the register values are restored from the values stored
on stack. This leads to an undesirable change in register values when transferring
control between two gadgets. Thus, the SPARC architecture and its calling
convention impose restrictions on the way the calculated values are passed between
gadgets~-- only through memory. The gadget catalog of Buchanan et al.~\cite{buchanan08} implements a
set of gadgets that use the memory-memory model, which allows the use of registers only
inside gadgets, and the values are passed from one gadget to another through memory.
Each variable in the ROP chain is associated with the address of the memory cell, which is
used as the gadget operand.

After the gadget cataloging, there are two options to create ROP chains automatically.
Firstly, the gadget catalog has a C programming interface. It contains 13 functions that
allow one to create variables, assign values, and call functions (or make system
calls). With this program interface, one can write a program that automatically
generates a ROP chain using a gadget catalog. Secondly, Buchanan et al.~\cite{buchanan08} wrote a
translator from some pseudo-language of the exploit description (narrowed C) into a
sequence of function calls from the gadget catalog program interface in C language. The
compiler implements most of the basic arithmetic, logical operations, operations with
pointers and memory, and operations of conditional and unconditional control transfer.

Some authors~\cite{buchanan08, ouyang15} note that it is possible to write an extension for the LLVM compiler
infrastructure, which allows one to generate code for the virtual machine defined by the
gadget catalog.

The tool, introduced by Mosier~\cite{roptools, mosier19}, relies on ROPC-IR, an exploit description
assembly-language, that defines a Turing-complete instruction set architecture. It has three
registers: \verb|ACC(eax)|, \verb|SP(rbp)|, \verb|PC(rsp)|, and operations for interacting with these
registers: basic arithmetic (\verb|ADD|, \verb|SUB|, \verb|NEG|), branch instructions (\verb|CMP|, \verb|JMP|, \verb|JNE|),
register-to-register (\verb|MOV|) instructions, register-to-memory (\verb|LD|, \verb|STO|) instructions, stack
instructions (\verb|PUSH|, \verb|POP|, \verb|ALLLOC|, \verb|LEAVE|), control transfer instructions (\verb|CALL|,
\verb|SYSCALL3|, \verb|LIBCALL3|). The \verb|rsp| register acts as the program counter (\verb|PC|). Moreover, a
separate stack is allocated for the functions inside the ROP chain, the pointer to which
\verb|SP| is \verb|rpb|.

Support for the second stack allows one to implement function calls inside the ROP
chain, which implement full-fledged subprograms. Besides, it supports the ability to
call functions from the target program address space and the ability to make
system calls. As an illustration, Mosier~\cite{roptools} gives an example of ROPC-IR code for
calculating Fibonacci numbers through a recursive function call from the ROP chain, the
\verb|printf| library function call, as well as the \verb|exit| system call.

ROPC-IR language description represents the gadget catalog in this tool.
The gadget search process prints all kinds of gadgets from the target program. Then it is
necessary to manually find and assign specific gadget to each semantic description and
manually catalog the following: virtual address, gadget
parameters. By definition of the ROPC-IR language gadgets have no side effects (not
taking into account output registers). Theoretically, the ROPC-IR assembler language
can act as the target C compiler language. However, the practical application for
building exploits for real programs can be significantly limited by the presence of the
necessary gadgets in the program and the size of the generated ROP chains. Due to
the nonoptimality of the ROPC-IR language translation, the chain size is
significantly larger than the typical exploit sizes.

The approach to building an automated tool for generating ROP chains, proposed in~\cite{buchanan08, roemer12, mosier19},
is based on a fixed gadget catalog. It is once formed by the authors and does not
change. Besides, semantic descriptions are tightly bound to the specific registers used in
gadgets. If some version of the standard \verb|libc| library lacks any gadget, then
ROP chain compilation may fail. In this case, it would be possible to use other gadgets that
have different operands, but similar functionality, and achieve successful compilation.
In other words, this approach has limited practical applicability, especially in situations
of a small number of gadgets in the library code under study.

%%%%%%%%%%%%%%%%%%%%%%%%%%%%%%%%%%%%%%%%
\subsection{Generation Based on Gadget Templates}
%%%%%%%%%%%%%%%%%%%%%%%%%%%%%%%%%%%%%%%%

Generation based on gadget templates is a search by regular expressions for a
specific sequence of gadgets that performs some malicious payload: the \verb|execve| system
call~\cite{ropgadget, ropper}, the \verb|VirtualProtect| function call, followed by the execution of a regular
shellcode on stack (Section~\ref{sec:virtualprotect})~\cite{monapy}, etc. Such an approach can handle
restricted symbols by negating the value loaded from the stack, by
repeating the increment
up to the desired value, or by other
arithmetic operations. It is worth noticing that Ropper~\cite{ropper} uses SMT solvers to
search for gadgets that satisfy the semantics defined by the postcondition
over registers, memory, and constants. However, at the time of writing this paper,
the tool uses only regular expressions to generate ROP chains.

Huang et al.~\cite{huang12}, for the ARM architecture, use an approach based on a unique gadget which
simultaneously loads the values of all registers from the stack. The searching algorithm
and simultaneous checking of the gadget for compliance with the given semantics are
performed by analyzing the assembler code instructions. Generating a chain from one
gadget is a trivial task. It requires only the correct location of the
register values on stack.

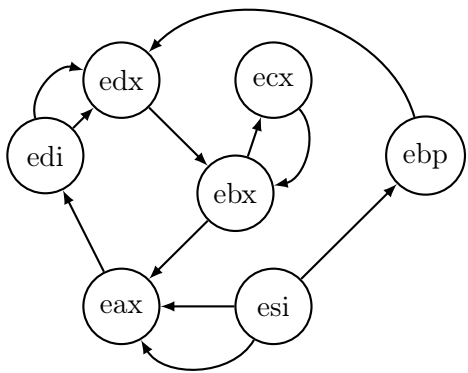
\begin{figure}[t]
    \centering
\tikzset{
    % Define standard arrow tip
    >=latex,
    % Define arrow style
    ptr/.style={->, thick},
    % Define gadget tree node style
    gad/.style = {shape=rectangle, rounded corners, draw, align=center, thick},
    opgad/.style = {gad, dashed},
    % Define gadget tree edge style
    edge from parent/.style = {draw, ptr},
    edgetoparent/.style = {draw, -, thick},
    % Define gadget tree dependency node
    dep/.style = {circle,draw,thick,minimum size=5mm},
}
    \begin{tikzpicture}
      [
        every node/.style = {circle,draw,thick,minimum size=1cm},
        every edge/.style = {draw, ptr},
        sloped
      ]
      \node(eax) at (0,0) {eax};
      \node(ebx) at (1.5,1.5) {ebx};
      \node(ecx) at (2,3) {ecx};
      \node(edx) at (0,3) {edx};
      \node(ebp) at (4,2) {ebp};
      \node(esi) at (2,0) {esi};
      \node(edi) at (-1,2) {edi};
      \path (edi) edge [bend left=60] (edx);
      \path (esi) edge [bend left=60] (eax);
      \path (ebp) edge [bend right=60] (edx);
      \path (ecx) edge [bend left=60] (ebx);
      \path (esi) edge (ebp);
      \path (eax) edge (edi);
      \path (ebx) edge (ecx);
      \path (edi) edge (edx);
      \path (edx) edge (ebx);
      \path (ebx) edge (eax);
      \path (esi) edge (eax);
    \end{tikzpicture}
    \caption{MOV connection graph. Chained gadgets can be used to emulate missing gadgets}
    \label{fig:move_chain_graph}
\end{figure}

Hund et al.~\cite{hund09} present another approach to gadget cataloging and compilation.
Firstly, they search only for gadgets consisting of one
instruction, not counting the return instruction itself. Most likely, it is done in
order to simplify the algorithms for analyzing the gadget parameters and side effects.
Such gadgets are cataloged. Secondly, they supplement the gadget
catalog with the gadgets that can be combined from existing ones.
The following example can illustrate it:
\begin{enumerate}
  \item \texttt{pop eax ; ret}~-- gadget loading value from stack to register \verb|eax|,
  \item \texttt{mov ebx, eax ; ret}~-- gadget moving the value from the register \verb|eax| to \verb|ebx|.
\end{enumerate}
These two gadgets, called sequentially, form a gadget for loading value from the stack
into the \verb|ebx| register. Hund et al.~\cite{hund09} provide an algorithm to search for all possible
combinations of gadgets that move a value from one register to another. This
task reduces to finding a path from one vertex to another in a special graph. In that
graph, registers act as vertices whereas initial gadgets, that move one register to
another, act as edges. Figure~\ref{fig:move_chain_graph} presents an example of such graph.

This approach allows one to expand the gadget catalog, which is especially useful for
exploitable programs of small size. However, using combined gadgets, one should
necessarily consider side effects at the stage of chaining gadgets.

Nguyen An Quin~\cite{quynh13} proposed a similar idea to combine several gadgets into one that
performs the desired behavior. For example, a gadget sequence
\texttt{push 0x1234 ; pop ebp ; ret ; xchg ebp, eax ; ret}
can be considered as one gadget loading the constant 0x1234 into the \verb|eax| register.

%%%%%%%%%%%%%%%%%%%%%%%%%%%%%%%%%%%%%%%%
\subsection{Chaining Gadgets by Semantic Queries}
\label{sec:ropium}
%%%%%%%%%%%%%%%%%%%%%%%%%%%%%%%%%%%%%%%%

Milanov~\cite{milanov18} obtains a semantic description of each gadget by building its
dependency graph (Section~\ref{sec:gadget-graph}). Gadgets are chained together by sequential
semantic queries to the gadget catalog. This method is implemented as an open-source
tool
ROPium~\cite{ropgenerator}. A semantic query is essentially an expression over constants, final/initial
values of registers and memory. First, gadgets with a semantic description that satisfy
the semantic query are searched in the gadget catalog. If such gadgets are missing, then
the semantic request is split into several ones according to some strategies. For example,
the first register can be moved to the second through some intermediate third register.

A notable feature of the ROPium tool is support of gadgets ending in
\texttt{call Reg} and \texttt{jmp Reg}
instructions. A load gadget for \texttt{Reg} register is
added before such gadgets. The gadget loads the gadget address to which it is necessary to
transfer control after executing the gadget with \texttt{call} or \texttt{jmp}.
In the case of \texttt{call}, it may
also be necessary to transfer control to a special gadget that removes the return address
placed by \texttt{call} from the stack.

%%%%%%%%%%%%%%%%%%%%%%%%%%%%%%%%%%%%%%%%
\subsection{Genetic Algorithm}
%%%%%%%%%%%%%%%%%%%%%%%%%%%%%%%%%%%%%%%%

Fraser et al.~\cite{fraser17, fraser18} suggest a different approach to building ROP chains. The authors
suggest using genetic algorithms for this. Fraser et al. introduced the ROPER tool
available on GitHub~\cite{roper}. The tool allows one to generate a ROP chain for the ARM
architecture, which sets registers to the specified values.

Initially, gadgets are searched in the executable. For each of them, the frame size and
the offset of the next gadget address are calculated (Section~\ref{sec:gadget-frame}). Then, the executable
file is loaded into the virtual machine address space for repeated ROP candidate chains execution. The virtual machine provides a convenient interface for
executing guest architecture instructions.

During genetic mutations, gadget addresses play the role of genes. Random
values are placed onto the stack as data and next gadget addresses. The fitness
function is the difference between the current and the required vector of register values. 
Genetic mutation methods modify each element of the population.
Among all the candidates, we select a set of potentially best candidates for
which the mutation process is repeated.

It is worth noticing that the chains formed
by the ROP genetic algorithm are very different from those created by people. For
example, they can write values to their stack or transfer control to gadgets that were not
originally found during the search process. Apart from that, the chain size can be
large due to the suboptimal choice of gadgets. The described disadvantages may come
from the lack of information about gadget instructions semantics in the genetic
algorithm. Perhaps, if we consider it in some form and use more modern
machine learning methods, we can develop the concept of this approach into a
practically useful tool.

%%%%%%%%%%%%%%%%%%%%%%%%%%%%%%%%%%%%%%%%
\subsection{Chain Generation by SMT-Solvers}
%%%%%%%%%%%%%%%%%%%%%%%%%%%%%%%%%%%%%%%%

Follner et al.~\cite{follner16} proposed a generation method based on a gadget
summary (Section~\ref{sec:summary}), which has an available source code~\cite{pshape}. The method allows one to get a sequence
of gadgets that writes the specified values into $m$ requested registers. It is worth noticing
that the method does not calculate the gadget parameters loaded from the stack, but only
provides a sequence of gadget addresses. Initially, a summary is compiled for all gadgets.
For each requested register, the algorithm selects $n$ most suitable gadgets~\cite{follner16_2}, which load the
register value from the stack or memory controlled by the attacker.

Further, for various gadget chains ($n^m * m!$ combinations), preconditions and
postconditions are calculated for the entire chain. If the postconditions satisfy
the situation when the attacker controls values of all requested registers, then the
method turns to the final stage~-- the preconditions resolving. Additional gadgets are
prepended to the beginning of the chain. These gadgets initialize registers from the
preconditions,
so that they point to the memory available for reading and writing.

\begin{algorithm}[t]
\caption{Search algorithm of shortest chain initializing registers.}
\small
\begin{algorithmic}
\State $regset\_to\_chain \gets$ empty register set mapping to shortest chains
\State $queue \gets$ empty queue
\State $queue.push($empty chain$)$
\While{$queue$ is not empty}
  \State $chain \gets queue.pop()$
  \ForAll{$gadget \in gadgets$}
    \State $new\_chain \gets chain + gadget$
    \State $regset \gets controlled\_registers(new\_chain)$
    \If{$regset$ not in $regset\_to\_chain$ or $new\_chain$ is shorter than $regset\_to\_chain[regset]$}
      \State $regset\_to\_chain[regset] \gets new\_chain$
      \State $queue.push(new\_chain)$
    \EndIf
  \EndFor
\EndWhile
\end{algorithmic}
\label{alg:angrop}
\end{algorithm}

Salls~\cite{angrop} developed the
method described above. Below, we describe a method for generating chain that sets register
values to specified values. The rest of chains, such as writing to
memory and calling a function, can be obtained by appending just one gadget to the chain
that initializes the registers. The method can be divided into three steps:
\begin{enumerate}
  \item \textbf{Gadget summary composing.} Symbolic execution~\cite{king76,
    schwartz10, godefroid08} of each gadget instructions occurs. Gadget
    summaries are composed through a static analysis of the resulting SMT
    expressions and queries to the SMT solver.
  \item \textbf{Chaining gadgets.} At this step, one searches for the shortest
    chains to initialize arbitrary sets of registers.
    Dijkstra's algorithm~\cite{dijkstra1959} inspired the proposed
    algorithm~\ref{alg:angrop} for finding the shortest paths from one of the
    graph vertices to all the others.
    An empty mapping is created from the
    register sets to the shortest chains that initialize these registers. An
    empty chain is added to the queue. The algorithm takes the chains from the
    queue. For each gadget, a new chain is created by appending this gadget to the
    chain taken from the queue. The set of registers initialized by a new chain
    ($controlled\_registers$) is calculated. If there is no such a set in the
    mapping, or the resulting chain is shorter than the one in the mapping, then
    a new chain is added into the mapping for this set. Also, the same chain is
    added to the queue. Thus, mapping is obtained from the sets of
    registers into the shortest chains that initialize these registers.
  \item \textbf{Placing a ROP chain on stack.} Symbolic execution of the
    entire ROP chain starts. Free symbolic variables are created for values
    loaded from the stack. At the end of the symbolic execution process, a
    conjunction of the equalities of the requested registers to the given values
    is constructed. As a result of solving this conjunction, the SMT solver
    provides bytes that are to be placed on stack.
\end{enumerate}
The described method, unlike the previous one, allows one to use those
gadgets in chains that initialize several registers at once, as well as gadgets that
perform arithmetic operations on registers loaded by other gadgets (SMT solver
calculates the correct values on stack). Moreover, this method allows for the shortest
chains to be selected.

Similar to the method of Follner et al.~\cite{follner16}, chains are generated by an open-source tool
Exrop~\cite{exrop} that builds gadget summaries as a result of their instructions symbolic
execution through the Triton framework~\cite{saudel15}. Suitable gadgets are selected for each
register to be set. An SMT solver checks postcondition satisfiability. It is worth noticing
that the tool supports jump-oriented (JOP) gadgets similar to ROPium (Section~\ref{sec:ropium}).

%%%%%%%%%%%%%%%%%%%%%%%%%%%%%%%%%%%%%%%%
\subsection{Semantic Tree Generation}
%%%%%%%%%%%%%%%%%%%%%%%%%%%%%%%%%%%%%%%%

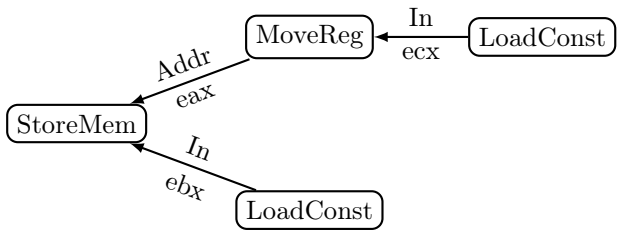
\begin{figure}[t]
    \centering
    \begin{tikzpicture}
      [
        grow              = right,
        sibling distance  = 6em,
        level distance    = 8em,
        every node/.style = {font=\small},
        sloped
      ]
      \tikzset{
          % Define standard arrow tip
          >=latex,
          % Define gadget tree node style
          gad/.style = {shape=rectangle, rounded corners, draw, align=center, thick},
          % Define gadget tree edge style
          edge from parent/.style = {draw, <-, thick},
          edgetoparent/.style = {draw, -, thick},
      }
      \node [gad] {StoreMem}
        child { node [gad] {LoadConst}
          edge from parent node [above] {In}
                           node [below] {ebx} }
        child { node [gad] {MoveReg}
          child { node [gad] {LoadConst}
            edge from parent node [above] {In}
                             node [below] {ecx} }
          edge from parent node [above] {Addr}
                           node [below] {eax} };
    \end{tikzpicture}
  \caption{Gadget tree that stores arbitrary value at arbitrary memory address}
  \label{fig:gadget-tree}
\end{figure}

Schwartz et al.~\cite{schwartz11} propose an approach to the ROP chains generation
based on semantic trees. The authors created the QooL language for
writing ROP chains, which is not Turing-complete but allows expressing ROP
chains used in practice (a library function call, system call, and writing to
memory). The process of translating a QooL program into a ROP chain consists of the
following steps:
\begin{enumerate}
  \item Generation of semantic trees by tiling~\cite{aho2006} the original QooL program abstract
    syntax tree. The semantic tree consists of
    abstract gadgets (gadget types) that define an instruction set architecture
    and are described in Section~\ref{sec:gadget-types}.
  \item Real gadgets (found in the program) assignment to abstract gadgets from
    the semantic tree. An example of a real gadgets tree is given in
    Figure~\ref{fig:gadget-tree}. Gadget types are stored in tree nodes. Type
    parameter names and their values (specific registers) are stored on
    edges. The gadget tree writes an arbitrary value to an arbitrary memory
    address. The stored value and the address are loaded from the stack
    into the $ebx$ and $ecx$ registers, respectively. The address from the $ecx$
    register is moved to the $eax$ register. After that, the value of the $ebx$
    register is stored at address $eax$.
  \item Scheduling a gadgets tree and generating a ROP chain.
\end{enumerate}

\begin{figure}[t]
  \centering
  \begin{tikzpicture}
    [
      every node/.style = {circle,draw,thick,minimum size=1cm},
      every edge/.style = {draw, ->, >=latex, thick},
      node distance     = 2cm,
      auto
    ]
    \node(a) {$a$};
    \node(c) [right of=a] {$c$};
    \node(b) [right of=c] {$b$};
    \path (a) edge [bend left=30] (b);
    \path (c) edge (b);
  \end{tikzpicture}
  \caption{Scheduling gadget tree}
  \label{fig:rop-schedule}
\end{figure}
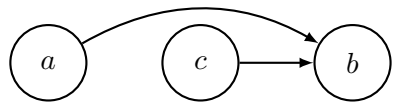

The first step is the lazy generation of all possible semantic trees from abstract gadgets.
It is necessary because some gadgets may not be available in a particular program. In
the second step, gadgets are assigned to each semantic tree. If it is not possible to assign
a specific gadget to each abstract gadget, then the semantic tree is discarded, and
the following is taken. In case of successful assignment, the tree of real
gadgets is passed to the third step. To generate the ROP chain, the gadget
tree must be linearized, i.e., scheduled. Gadgets tree scheduling should
consider: data dependencies between the gadget registers and ``clobbered''
registers. It means the following (Fig.~\ref{fig:rop-schedule}):
\begin{enumerate}
  \item The schedule should satisfy the topological order of the tree.
  \item If the output register of gadget $a$ is used by gadget $b$, then this
    register should not be ``clobbered'' by any gadget in the schedule between $a$
    and $b$.
\end{enumerate}

During semantic trees generation, certain gadget
types absence possibility is considered, and all available rules are applied sequentially to
express the vertices of the abstract syntax tree through semantic trees
from abstract gadgets. For example, the authors noticed that the ROP chain
generation success increases if the following rule for expressing a vertex, that
stores
a value in memory, is added:
\begin{enumerate}
  \item \texttt{mov [eax], 0 ; ret}
  \item \texttt{pop ebx ; ret}
  \item \texttt{add [eax], ebx ; ret}
\end{enumerate}

Ouyang et al.~\cite{ouyang15} extended the QooL instruction set to a Turing-complete set. In
general, they repeat the approach of Schwartz et al.~\cite{schwartz11} with the construction of
semantic trees, using a value liveness analysis when dealing with side effects.
It is worth noticing that there are attempts to implement the method of Schwartz et al. that
have open-source code~\cite{ropc, pyrop, pyrop-paper}.

%%%%%%%%%%%%%%%%%%%%%%%%%%%%%%%%%%%%%%%%
\section{RESTRICTED SYMBOLS}
\label{sec:badchars}
%%%%%%%%%%%%%%%%%%%%%%%%%%%%%%%%%%%%%%%%

Automatic ROP chain generation tools should deal with input sanitization for a
particular exploit. For example, data copied through \texttt{strcpy} function cannot contain
null bytes. Both gadget addresses and values loaded by gadgets from the stack can
contain restricted symbols.

In simplest case, gadget addresses are sanitized by dropping gadgets that contain
restricted symbol in the address. Many tools act the same. However, this approach inevitably
leads to a reduction in the gadget catalog, which causes a lack of gadgets and the need
to combine them to model missing gadgets.

The situation is much more complicated when restricted symbols are in data to be
loaded onto registers (values of function arguments and values necessary for writing
to memory). To solve this problem, one can use various arithmetic operations to obtain
values containing restricted symbols.

A detailed description of dealing with restricted symbols is given in an article by
Ding et al~\cite{ding14}. It is worth noting that the authors are restricting all non-printable
characters, which may be excessive in some cases. However, their methods are
applicable in the more general case of an arbitrary set of restricted symbols. For each
gadget, the authors construct a semantic tree that describes the gadget functionality
and contains explicit dependencies between registers and memory regarding
arithmetic operations and memory interaction operations.

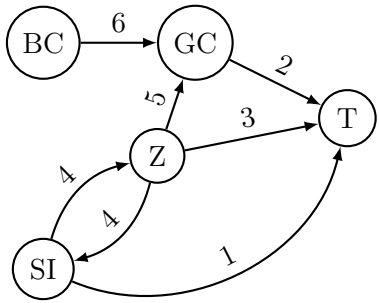
\begin{figure}[t]
  %{{{
 \centering
\tikzset{
  >=latex,
  c/.style = {shape=circle,draw,thick,align=center,minimum width=0.5cm},
  g/.style = {align=left}
}
  \begin{tikzpicture}
    [
      every edge/.style = {draw, ->, thick},
      sloped
    ]
    \node[c](SI) at (0,0) {SI};
    \node[c](Z) at (1.5,1.5) {Z};
    \node[c](GC) at (2,3) {GC};
    \node[c](BC) at (0,3) {BC};
    \node[c](T) at (4,2) {T};
    \path (SI) edge[bend left=30] node[g, midway, above] {4} (Z);
    \path (Z) edge[bend left=30] node[g, midway, above] {4} (SI);
    \path (Z) edge node[g, midway, above] {5} (GC);
    \path (BC) edge node[g, midway, above] {6} (GC);
    \path (GC) edge node[g, midway, above] {2} (T);
    \path (Z) edge node[g, midway, above] {3} (T);
    \path (SI) edge[bend right=50]  node[g, midway, above] {1} (T);
  \end{tikzpicture}
  \caption{State machine describing the input sanitizing algorithm}
  \label{fig:ding_state_machine}
  %}}}
\end{figure}

The constructed semantic trees help to build a finite state machine
(Fig.~\ref{fig:ding_state_machine}). This FSM is intended to search for instructions that load a value into a register. The vertices in
the finite state machine are the following states corresponding to different loaded
values:
\begin{itemize}
  \item Z~-- zero,
  \item SI~-- small number,
  \item GC~-- number containing no restricted symbols,
  \item BC~-- number containing restricted symbols,
  \item T~-- final state.
\end{itemize}

Between these vertices, there are edges corresponding to specific gadgets,
provided by the gadget catalog. Possible transitions between states of a finite state
machine are the following:
\begin{enumerate}
  \item $SI \longrightarrow T$, the edge from the vertex with a small number to
    the final state corresponds to the gadget with the instruction directly
    initializing this value in the register.
  \item $GC \longrightarrow T$, an edge with a \texttt{pop} gadget leads from a
    vertex with a number that does not contain restricted symbols to the final
    state.
  \item $Z \longrightarrow T$, an edge with xor instruction leads from a vertex
    with zero to the final state.
  \item $SI \longleftrightarrow Z$, edges with instructions \texttt{inc},
    \texttt{dec} lead from number to zero and vice versa
  \item $Z \longrightarrow GC$, an edge with arithmetic instructions
    \texttt{and}, \texttt{or}, \texttt{sal}, \texttt{shl}, \texttt{shr},
    \texttt{sar} leads from the vertex with zero to a state with a number that
    does not contain restricted symbols.
  \item $BC \longrightarrow GC$, edges consisting of a combination of two
    arithmetic operations, for example, $a + b - c$, lead from a vertex with a
    number containing restricted symbols to a vertex that does not contain
    restricted symbols.
\end{enumerate}

The algorithm starts from state corresponding to the value that needs to be set in
a particular register. By traversing the states of this state machine, one can solve
the problem of possible ROP chain data sanitization. The algorithm is interrupted
if (1) the final state is reached, which corresponds to the successful finding of a
combination of gadgets that solve the task, or (2) there are no transitions to other
states from the current one.

%%%%%%%%%%%%%%%%%%%%%%%%%%%%%%%%%%%%%%%%
\section{EXPERIMENTAL TOOLS COMPARISON}
\label{sec:evaluation}
%%%%%%%%%%%%%%%%%%%%%%%%%%%%%%%%%%%%%%%%

\begin{table*}[t]
\caption{The experimental evaluation of automatic ROP chain generation tools}
\centering
\scriptsize
\begin{tabular}{ l | >{\columncolor[gray]{0.9}}r r r | >{\columncolor[gray]{0.9}}r r r | >{\columncolor[gray]{0.9}}r r r | >{\columncolor[gray]{0.9}}r r r | >{\columncolor[gray]{0.9}}r r r }
\toprule
Test suite & \multicolumn{3}{c |}{Synthetic} & \multicolumn{3}{c |}{OpenBSD 6.4}
           & \multicolumn{3}{c |}{OpenBSD 6.2} & \multicolumn{3}{c |}{Debian 10}
           & \multicolumn{3}{c}{CentOS 7} \\
Number of files & \multicolumn{3}{c |}{22} & \multicolumn{3}{c |}{410} & \multicolumn{3}{c |}{397} & \multicolumn{3}{c |}{689} & \multicolumn{3}{c}{649} \\
Has syscall gadget & \multicolumn{3}{c |}{21} & \multicolumn{3}{c |}{98} & \multicolumn{3}{c |}{87} & \multicolumn{3}{c |}{139} & \multicolumn{3}{c}{121} \\
At least one OK & \multicolumn{3}{c |}{13} & \multicolumn{3}{c |}{19} & \multicolumn{3}{c |}{50} & \multicolumn{3}{c |}{115} & \multicolumn{3}{c}{72} \\
\midrule
 Tool         &   OK &   F &   TL &   OK &   F &   TL &   OK &   F &   TL &   OK &   F &   TL &   OK &   F &   TL \\
 \href{https://github.com/JonathanSalwan/ROPgadget/tree/c29c50773ec7fb3df56396ce27fb71c3898c53ae}{ROPgadget}~\cite{ropgadget}
              &    1 &   0 &    0 &    2 &   0 &    0 &    4 &   0 &    0 &    7 &   0 &    0 &    8 &   0 &    0 \\
 \href{https://github.com/sashs/Ropper/tree/75a9504683427e373c7bb6d6a54ed20bd98905ff}{Ropper}~\cite{ropper}
              &    2 &  -- &    0 &    3 &  -- &    0 &   15 &  -- &    0 &   53 &  -- &    0 &   31 &  -- &    0 \\
 \href{https://github.com/d4em0n/exrop/tree/343eee05bd4b9d31db3e55a70a33893527225c84}{Exrop}~\cite{exrop}
              &    9 &   0 &    0 &    0 &  33 &   28 &   11 &  27 &   13 &   76 &  19 &    5 &   48 &   8 &   12 \\
 \href{https://github.com/salls/angrop/tree/794583f59282f45505a734b21b30b982fceee68b}{angrop}~\cite{angrop}
              &    8 &   1 &    0 &   10 &   1 &    2 &   25 &   2 &    3 &   86 &  12 &    1 &   54 &   9 &    0 \\
 \href{https://github.com/Boyan-MILANOV/ropium/tree/e7100878b75e55d775eecfd79bd549f9895f4c8c}{ROPium}~\cite{ropgenerator}
              &   10 &   1 &    0 &   18 &   4 &    0 &   43 &   6 &    1 &  103 &  10 &    0 &   64 &  11 &    0 \\
\bottomrule
\end{tabular}
\label{tbl:exp_results}
\end{table*}

We performed experimental testing of tools, that have an available source code, with
rop-benchmark~\cite{ropbenchmark} test system. This system provides a reproducible environment for
checking generation success and exploitability of ROP chains invoking
\texttt{execve("/bin/sh", 0, 0)} system call. The testing system supports Linux
x86-64 platform. We took executable files and libraries from minimal installations of several
popular distributions: CentOS 7, Debian 10, OpenBSD 6.2, OpenBSD 6.4. We
considered
OpenBSD 6.2 and 6.4 because the authors of this operating system intentionally
reduced the number of ROP gadgets~\cite{mortimer19}.

Table~\ref{tbl:exp_results} presents experimental evaluation results.
Four columns correspond to four sets of test files. The first line shows the
total number of test files in each set. The second line~-- the number of
binaries that contain syscall gadgets. The third line~-- the number of files
that have a working ROP chain created by at least one tool. Below are the
lines with the tools, and next to each tool the following information is
indicated:
\begin{itemize}
  \item OK~-- the number of test files for which the created ROP chain is
    exploitable, i.e., leads to the opening of the system shell.
  \item F~-- the number of test files for which the created ROP chain is not
    workable, i.e., for some reason, it does not open the system shell. It is
    worth noticing that we performed 10 runs of an executable file. If at least
    one did not lead to shell opening, the generated chain is not considered
    workable.
  \item TL~-- the number of test files on which the tool runtime exceeded the
    set limit of 1 hour.
\end{itemize}
Ropper almost always generates a ROP chain script file, so the F number was not
evaluated.

The experimental comparison included only publicly available tools that could
fully automatically generate a ROP chain which performs a system call for
x86-64 architecture on Linux family operating system. We did not consider the
mona.py tool~\cite{monapy} due to the operating system. Others can work only
with x86 architecture (32-bit)~\cite{sqlabropchain}, ARM~\cite{roper}. Some
available tools failed to be successfully integrated into the automated
execution system~\cite{ropc}. The test system does not include tests with
restricted symbols (for instance, 0x00 in case of overflow while copying via
\texttt{strcpy}) because no tool completely supports them,
i.e., checks the presence of restricted symbols not only in the gadget addresses
but also in gadget parameters values on stack.
Tools ROPium~\cite{ropgenerator}, Ropper~\cite{ropper} can only restrict
symbols in gadget addresses.

%%%%%%%%%%%%%%%%%%%%%%%%%%%%%%%%%%%%%%%%
\section{CONCLUSIONS}
\label{sec:conclusion}
%%%%%%%%%%%%%%%%%%%%%%%%%%%%%%%%%%%%%%%%

This article provides a detailed overview of code reuse attacks and
automated exploit generation techniques for such attacks. Code reuse attacks
imply the use of code pieces from the program address space, the pieces are called
\textit{gadgets}. Gadgets are linked in a chain that performs a malicious payload. We divide
the process of code reuse exploits generation into four stages:
searching for gadgets in an exploitable program, determining gadgets semantics,
combining gadgets in chains, and generating input data exploiting the vulnerability. In
the first stage, found gadgets constitute a gadget catalog. After that, one derives the
gadgets semantics and catalog them. There are three ways to present gadget
semantics: parameterized semantic types, gadget summaries, gadget
dependency graphs. In the third stage, gadgets can be chained both by searching
according to regular expression templates or considering their semantics.

In some
cases, if the set of gadgets in the catalog is Turing-complete, then the gadgets can be
used as the target architecture for a compiler. Moreover, some approaches construct
ROP chains with genetic algorithms, while others use SMT solvers.
It is worth noting that methods for automated generation of chains
that exploit data flow (DOP) are beyond the article.

We propose the ROP Benchmark~\cite{ropbenchmark} for experimental comparison of ROP chain
generation tools. We compared open-source ROP chain generation tools on
Linux x86-64 platform. In particular, the comparison was carried out on OpenBSD
distributions, the authors of which intentionally reduce the number of ROP gadgets~\cite{mortimer19}.

During chains generation, it is crucial to take account of restricted symbols. For
example, if \texttt{strcpy} function processes input data, they cannot contain zero bytes.
However, just a few authors consider restricted symbols in the chain generation methods.

There are a wide range of code-reuse attack methods (ROP, JOP, and others). The
question of what set of such methods appears to be enough to implement exploitation
is still open. For example, an attacker could manually form a JOP chain for some
vulnerable program, while advanced methods allowed generating a regular ROP chain
for the same program.
The question arises whether it is possible to improve chain generation methods
without use of
complicated code-reuse attacks.

A promising direction for research is investigating how to bypass address space
randomizing protections without information leakage and brute-force.

Most methods
do not use gadgets that have arbitrary memory dereference as a side-effect~\cite{follner16}. Their
consideration would expand the gadget catalog and improve chain generation methods.

%%%%%%%%%%%%%%%%%%%%%%%%%%%%%%%%%%%%%%%%
\section*{FUNDING}
\label{sec:funding}
%%%%%%%%%%%%%%%%%%%%%%%%%%%%%%%%%%%%%%%%

This work was supported by the Russian Foundation for Basic Research, project
no. 17-01-00600.

\phantomsection
\printbibliography

\hfill \textit{Translated by M.~Zhizhchenko}

\label{lastpage}
\end{document}